\documentclass[12pt]{article} 
\usepackage{epsfig}
\usepackage{a4}
\usepackage{latexsym}
\usepackage{cite}

\usepackage{pslatex}
\usepackage[latin1]{inputenc}
\usepackage[T1]{fontenc}

\usepackage{color,colordvi}
\def\colour4colour#1{\Blue{#1}}

\newcommand{\gsim}{\raisebox{-0.07cm}{$\:\:\stackrel{>}{{\scriptstyle
 \sim}}\:\: $} }
\newcommand{\lsim}{\raisebox{-0.07cm}{$\:\:\stackrel{<}{{\scriptstyle
 \sim}}\:\: $} }
\newcommand{\hspn}{{\hspace{-7mm}}}
\newcommand{\hspp}{{\hspace{5mm}}}
\newcommand{\beq}{\begin{equation}}
\newcommand{\eeq}{\end{equation}}
\newcommand{\bea}{\begin{eqnarray}}
\newcommand{\eea}{\end{eqnarray}}
\newcommand{\nn}{\nonumber}
\newcommand{\MSb}{$\overline{\mbox{MS}}$}
\newcommand{\as}{\alpha_{\rm s}}
\newcommand{\ar}{a_{\rm s}}
\newcommand{\ra}{\rightarrow}
\newcommand{\ep}{\epsilon}
\newcommand{\lam}{\lambda}
\newcommand{\GE}{\gamma_{\rm e}}
\newcommand{\GEs}{\gamma_{\rm e}^{\:2}}

\begin{document}
\setlength{\parskip}{0.15cm}
\setlength{\baselineskip}{0.53cm}

\def\Fone{{F_{\:\! 1}}}
\def\Ftwo{{F_{\:\! 2}}}
\def\FL{{F_{\:\! L}}}
\def\F3{{F_{\:\! 3}}}
\def\Qs{{Q^{\, 2}}}
\def\GeV2{{\mbox{GeV}^{\:\!2}}}
\def\DDk{{D}_{\:\!k}}
\def\x1{{(1 \! - \! x)}}
\def\DD#1{{{\cal D}_{\! #1}^{}}}
\def\z#1{{\zeta_{\:\! #1}}}
\def\zss{{\zeta_{2}^{\,2}}}
\def\zst{{\zeta_{3}^{\,2}}}
\def\zts{{\zeta_{2}^{\,3}}}
\def\ca{{C^{}_A}}
\def\cas{{C^{\: 2}_A}}
\def\cat{{C^{\: 3}_A}}
\def\cf{{C^{}_F}}
\def\cfs{{C^{\: 2}_F}}
\def\cft{{C^{\: 3}_F}}
\def\cff{{C^{\: 4}_F}}
\def\nf{{n^{}_{\! f}}}
\def\nfs{{n^{\,2}_{\! f}}}
\def\nft{{n^{\,3}_{\! f}}}
\def\dabc2{{d^{\:\!abc}d_{abc}}}
\def\dabcnc{{{d^{\:\!abc}d_{abc}}\over{n_c}}}
\def\fl11{fl_{11}}
\def\fl02{fl_{02}}
\def\b#1{{{\beta}_{#1}}}
\def\bb#1#2{{{\beta}_{#1}^{\,#2}}}

\def\S(#1){{{S}_{#1}}}
\def\Ss(#1,#2){{{S}_{#1,#2}}}
\def\Sss(#1,#2,#3){{{S}_{#1,#2,#3}}}
\def\Ssss(#1,#2,#3,#4){{{S}_{#1,#2,#3,#4}}}
\def\pqq(#1){p_{qq}(#1)}
\def\gfunct#1{{g}_{#1}^{}}

\def\H(#1){{\rm{H}}_{#1}}
\def\Hh(#1,#2){{\rm{H}}_{#1,#2}}
\def\THh(#1,#2){{\widetilde{\rm{H}}}_{#1,#2}}
\def\Hhh(#1,#2,#3){{\rm{H}}_{#1,#2,#3}}
\def\THhh(#1,#2,#3){{\widetilde{\rm{H}}}_{#1,#2,#3}}
\def\Hhhh(#1,#2,#3,#4){{\rm{H}}_{#1,#2,#3,#4}}
\def\Hhhhh(#1,#2,#3,#4,#5){{\rm{H}}_{#1,#2,#3,#4,#5}}

\begin{titlepage}
\noindent
DESY 09-133 \hfill {\tt arXiv:0909.2124 [hep-ph]}\\
SFB/CPP-09-80 \\
LTH 840 \\[1mm]
September 2009 \\
\vspace{1.5cm}
\begin{center}
\LARGE
{\bf On non-singlet physical evolution kernels and \\[1mm]
     large-{\em x} coefficient functions in perturbative QCD}

\vspace{2.5cm}
\large
S. Moch$^{\, a}$ and A. Vogt$^{\, b}$\\
\vspace{1.5cm}
\normalsize
{\it $^a$Deutsches Elektronensynchrotron DESY \\
\vspace{0.1cm}
Platanenallee 6, D--15738 Zeuthen, Germany}\\
\vspace{0.5cm}
{\it $^b$Department of Mathematical Sciences, University of Liverpool \\
\vspace{0.1cm}
Liverpool L69 3BX, United Kingdom}\\[1.5cm]
\vfill
\large
{\bf Abstract}
\vspace{-1mm}
\end{center}
We study the large-$x$ behaviour of the physical evolution kernels for 
flavour non-singlet observables in deep-inelastic scattering, where $x$ 
is the Bjorken variable, semi-inclusive $e^+e^-$ annihilation and Drell-Yan
lepton-pair production. Unlike the corresponding \MSb-scheme coefficient 
functions, all these kernels show a single-logarithmic large-$x$ enhancement 
at all orders in $1\!-\!x$. 
We conjecture that this universal behaviour, established by Feynman-diagram 
calculations up to the fourth order, holds at all orders in the strong coupling
constant $\as$. The resulting predictions are presented for the highest 
$\ln^{\,n}\!\x1$ contributions to the higher-order coefficient functions. 
In Mellin-$N$ space these predictions take the form of an exponentiation 
which, however, appears to be less powerful than the well-known soft-gluon 
exponentiation of the leading $\x1^{-1} \ln^{\,n}\!\x1$ terms.  In particular 
in deep-inelastic scattering the $1/N$ corrections are non-negligible for all 
practically relevant $N$.

\vfill
\end{titlepage}
%
%
\setcounter{equation}{0}
\section{Introduction}
\label{sec:intro}
%
%
Disregarding power corrections, hard hadron processes are described in 
perturbative QCD in terms of process-dependent short-distance coefficient 
functions (mass-factorized partonic cross sections) and universal space- and 
timelike parton densities including non-perturbative long-distance effects. The
separation between the coefficient functions and the parton densities and the
splitting functions governing their scale dependence is, of course, not unique 
beyond the leading order (LO) in perturbative QCD. It is usual to perform this 
separation in the modified~\cite{BBDM78} minimal subtraction \cite{MS} scheme, 
\MSb, see also Ref.~\cite{FP82}, of dimensional regularization \cite{DimReg}, 
the standard framework for higher-order diagrammatic calculations in quantum
field theory.

While this scheme does not provide a physical definition of the parton 
densities, it does lead to a stable (order-independent) functional form of the 
dominant diagonal (quark-quark and gluon-gluon) splitting functions in the
limit of large momentum fractions $x$ \cite
{Korchemsky:1989si,MVV3,MVV4,DMS05,MMV1,MV2}.
This feature assists a stable evolution of the parton densities over a wide 
range of scales as required, e.g., for LHC predictions based on data from 
fixed-target and HERA experiments. 

The coefficient functions, on the other hand, receive double logarithmic 
large-$x$ enhancements in the \MSb\ scheme, i.e., terms up to $\x1^{-1+k\,} 
\ln^{\,2n-a}\! \x1$ occur, for all $k$, at the $n$-th order of the strong 
coupling constant $\as$ (the offset $a \geq 1$ depends on the observable and 
the power $k$ in the expansion in $1\!-\!x\,$). The highest leading ($\,k=0\,$)
logarithms can be resummed by the soft-gluon exponentiation \cite{SoftGlue} 
which is now known to the next-to-next-to-next-to-leading logarithmic (N$^3$LL) 
accuracy for inclusive deep-inelastic lepton-proton scattering (DIS), 
$lp\ra l + X$ \cite{MVV7}, Drell-Yan (DY) lepton-pair production and Higgs 
production in proton-proton collisions \cite{MV1,Laenen:2005uz,Idilbi:2005ni}, 
and semi-inclusive electron-positron annihilation (SIA), 
$e^+e^- \!\ra\, h + X\,$ ($\,h = \pi,\, K,\, \dots$) \cite{Blumlein:2006pj,MV4}.
On the other hand, recent studies of the subleading $k=1$ logarithms 
\cite{OneoverN,Laenen:2008gt} have not led to similarly systematic predictions 
for higher-order coefficient functions yet.

An alternative description of hard processes can be obtained by eliminating the
parton densities, leading to physical evolution kernels (also called physical 
anomalous dimensions) for the scale dependence of observables~\cite{FP82}, see
also Ref.~\cite{Blumlein:2000wh}. This is especially simple for the flavour 
non-singlet quantities dominating the large-$x$ limits of the semi-leptonic
DIS, SIA and DY processes mentioned above.
The direct relation between 
different processes via the universal parton densities is absent in this 
approach, but the soft-gluon exponentiation \cite{SoftGlue} guarantees an only 
single logarithmic $k=0$ higher-order large-$x$ enhancement \cite{NV3}, see 
also Refs.~\cite{FR05GR05}.

Using the coefficient-function results of Refs.~\cite{ZvNcq2,Moch:1999eb,MVV2,%
MVV5,MVV6,MVV10,RvN96,Mitov:2006wy,Hamberg:1991np,Harlander:2002wh} for the 
above processes, one finds that the corresponding non-singlet physical kernels 
exhibit only a single logarithmic enhancement for all values of $k$ at least to
the next-to-next-to-leading or next-to-next-to-next-to-leading order (NNLO or 
N$^{\:\!3}$LO) in the expansion in $\as$. 
We are thus led to the rather obvious conjecture that this behaviour, already 
established to order $\as^{\,4}$ in DIS, persists to all orders in $\as$. The 
required cancellation of double logarithms in the physical kernels then implies
exponentiations also of terms with $k \geq 1$, yielding explicit all-order 
predictions for the highest logarithms in the respective quark coefficient 
functions. In the rest of this article we derive and discuss these predictions, 
emphasizing the subleading $k=1$ logarithms. Especially for this case a formal 
proof of the exponentiation may be expected in the near future from the new 
path-integral approach of Ref.~\cite{Laenen:2008gt}.
%
%
\setcounter{equation}{0}
\section{Physical evolution kernels for non-singlet observables}
\label{sec:general}
%
%
We start by recalling the construction and fixed-order properties of the
physical evolution kernels. We first consider the DIS structure functions
(see Ref.~\cite{PDG08} for a general overview)
\beq
\label{Fns-def}
  {\cal F}_{1}^{} \:=\: 2F_{\:\!1,\rm ns}\:\: , \quad
  {\cal F}_{2}^{} \:=\: \frac{1}{x}\, F_{\:\!2,\rm ns}\:\: , \quad
  {\cal F}_{3}^{} \:=\: \F3^{\!\nu + \bar{\nu}} \:\: .
\eeq
The longitudinal structure function $\FL = \Ftwo - 2x \Fone$ has been addressed
already in Ref.~\cite{MV3}. Disregarding terms suppressed by powers of $1/\Qs$,
the non-singlet quantities (\ref{Fns-def}) are given by
\beq
\label{Fns-cq}
  {\cal F}_a(x,\Qs)
  \:\: = \; \left[ C_a(\Qs) \otimes q_{a,\rm ns}^{\,}(\Qs) \right] \! (x)
  \; =\:\: {\displaystyle \sum_{l=0} } \: a_s^{\, l}(\Qs) \:\left[ c_{a,l} 
      \otimes q_{a,\rm ns}^{\,}(\Qs) \right] \! (x) \:\: .
\eeq
As usual $x$ is the Bjorken variable, and $\Qs = -q^{\:\!2}$ the negative 
squared four-momentum of the exchanged gauge boson. $c_{a,l}$ represents the 
$l$-loop non-singlet coefficient function for ${\cal F}_a$ with 
\mbox{$c_{a,0}(x)=\delta(1\!-\!x)$}. 
The exact three-loop results $c_{a,3}(x)$ for the structure functions 
(\ref{Fns-def}) have been computed in Refs.~\cite {MVV5,MVV6,MVV10}. Beyond 
this order only the $C_F n_{\!f}^{\:l-1}$ leading-$\nf$ terms are exactly known
\cite{Gracey:1995aj,Mankiewicz:1997gz}. 
Furthermore $q_{a,\rm ns}$ denotes the corresponding combination of the quark 
densities (including electroweak charge factors), and $\otimes$ stands for the
standard Mellin convolution, given by
\beq
\label{Mconv}
  [ \:\! a \otimes b ](x) \;\; = \;\; \int_x^1 \! \frac{dy}{y} \: a(y)\:
  b\bigg(\frac{x}{y}\bigg)
\eeq
for two regular functions and Eq.~(3.4) of Ref.~\cite{NV3} if a 
$+$-distribution is involved. The renormalization and factorization scales 
$\mu_{\:\!\rm r\,}$ and $\mu_{\:\!\rm f\,}^{}$ have been set to the physical 
hard scale $Q$ in Eq.~(\ref{Fns-cq}).

The scale dependence of the running coupling of QCD, in this article 
normalized as 
$$
  a_s \;\equiv\; \frac{\alpha_s}{4\pi} \:\: , 
$$
is governed by 
\beq
\label{asrun}
  \frac{d a_s}{d \ln \Qs} \; = \; \beta(a_s) \; = \;
  - \sum_{l=0} a_s^{\, l+2} \,\beta_l \:\: .
\eeq
Besides the scheme-independent $\beta_0 \,=\, 11/3\;C_A - 2/3\;\nf$ 
\cite{beta0} (with $\,\ca = N_{\rm colours} = 3$ in QCD) and $\beta_1$ 
\cite{beta1}, also the coefficients $\beta_2$ and $\beta_3$ have been computed 
\cite{beta2,beta3} in the \MSb\ renormalization scheme adopted throughout this 
study. All these four coefficients are required for  calculations including the
N$^{\:\!3}$LO quantities $c_{a,3}$. Here and below $\nf$ denotes the number of 
effectively massless flavours (mass effects are not considered in this 
article). 
Finally the evolution equations for the quark densities in Eq.~(\ref{Fns-cq}) 
read
\bea
\label{qevol}
  \frac{d}{d \ln Q^2} \; q_{a,\rm ns}^{\,}(x, Q^2) 
  & =\, & \left[ P_a(Q^2) \otimes q_{a,\rm ns}^{\,}(Q^2)\right] \! (x)
  \nonumber \\ & =\, & 
        \sum_{l=0} \, a_s^{\, l+1} \left[ P_{a,l}\otimes 
        q_{a,\rm ns}^{\,}(Q^2)\right] \! (x) \:\: .
\eea
As the coefficient functions $c_{a,l\,}$, the $(l\! +\! 1)$-loop splitting 
functions $P_{a,l}$ depend only on $x$ for the above choice of 
$\mu_{\:\!\rm r}$ and $\mu_{\:\!\rm f\,}^{}$. All three independent third-order
(NNLO) non-singlet splitting functions $P_{a,2}(x)$ are known from 
Ref.~\cite{MVV3}.

The convolutions in Eqs.~(\ref{Fns-cq}) and (\ref{qevol}) correspond to simple
products of the respective Mellin transforms given by
\beq
\label{Ndef}
  a^{\,N} \; = \; \int_0^1 \! dx \; x^{\,N-1} a(x)
\eeq
for regular functions such as $q_{a,\rm ns\,}$ and 
\beq
\label{Ndef2}
  a^{\,N} \; = \; \int_0^1 \! dx \, \left( x^{\,N-1} - 1 \right) a(x)_+
\eeq
for +-distributions such as the leading large-$x$ contributions to 
$c_{a,l\,}(x)$. Hence calculations involving multiple convolutions as, e.g., 
in Eqs.~(\ref{ctilde}) below are best carried out in $N$-space where the 
coefficient functions and splitting functions are expressed in terms of
harmonic sums \cite{Hsums}. We mainly use {\sc Form} \cite{Vermaseren:2000nd} 
and {\sc TForm} \cite{Tentyukov:2007mu} to manipulate such expressions, to 
transform them back to the $x$-space harmonic polylogarithms 
\cite{Remiddi:1999ew}, see also Ref.~\cite{Moch:1999eb}, and to extract 
large-$x$ coefficients from the results.

It is convenient, both phenomenologically -- for instance for determinations 
$\as$ -- and theoretically, to express the scaling violations of non-singlet 
observables in terms of these observables themselves. This explicitly 
eliminates any dependence on the factorization scheme and the associated scale 
$\mu_{\:\!\rm f\,}^{}$, and avoids the non-negligible dependence of the 
\MSb-scheme initial distributions for $q_{a,\rm ns}$ on the perturbative order.
The corresponding physical evolution kernels $K_a$ can be derived for 
$\mu_{\:\!\rm r}^{\,2} = \Qs$ by differentiating Eq.~(\ref{Fns-cq}) with 
respect to $\Qs$ by means of the respective evolution equations (\ref{asrun}) 
and (\ref{qevol}) for $\,\ar$ and $q_{a,\rm ns}$, and then using the inverse of 
Eq.~(\ref{Fns-cq}) to eliminate $q_{a,\rm ns}$ from the result. 
This procedure yields the evolution equations \cite{NV3} 
\bea
\label{Fevol}
  \frac{d}{d \ln \Qs} \; {\cal F}_a 
  &\!\! =\! &
  \bigg\{ P_a(\ar) + \,\beta(\ar)\: \frac{d\, C_a(\ar)}{d \ar}\, 
   \otimes \, C_a(\ar)^{\,-1} \bigg\} \otimes \, {\cal F}_a
\nn\\[3mm]
  & \!\! \equiv \! & \;
  K_a \otimes\, {\cal F}_a  \; \equiv \;\;
    \sum_{l=0} \, \ar^{\, l+1}\, K_{a,l} \otimes \, {\cal F}_a
\nn\\[-0.5mm]
  & \!\! =\! &
  \Bigg\{ \ar\, P_{a,0} \,+\: \sum_{l=1} \ar^{\, l+1}
  \Bigg( P_{a,l} - \sum_{k=0}^{l-1} \,\beta_k \:
  \tilde{c}_{a,l-k} \Bigg) \!\Bigg\}
  \otimes \: {\cal F}_a \:\: .
\eea
Notice that in $N$-space the second term in the first line simply is
$\beta(\ar)\: d\ln\, C_a(\ar) /d \ar$. Up to N$^{\:\!4}$LO (terms up to $l=4$
included in the sums) the expansion coefficients $\tilde{c}_{a,l}(x)$ in the 
last line read
\bea
\label{ctilde}
  \tilde{c}_{a,1} &\, =\, & c_{a,1}
  \nn \\
  \tilde{c}_{a,2} &\, =\, & 2\, c_{a,2} - c_{a,1}^{\,\otimes 2}
  \nn \\
  \tilde{c}_{a,3} &\, =\, & 3\, c_{a,3} - 3\, c_{a,2} \otimes c_{a,1}
    + c_{a,1}^{\,\otimes 3}
  \\
  \tilde{c}_{a,4} &\, =\, & 4\, c_{a,4} - 4\, c_{a,3} \otimes c_{a,1}
    - 2\, c_{a,2}^{\,\otimes 2} + 4\, c_{a,2} \otimes
   c_{a,1}^{\,\otimes 2} - c_{a,1}^{\,\otimes 4} 
 \nn
\eea
with $ f^{\,\otimes 2} \equiv f \otimes f$ etc. 
The above expressions for $K_{a,l\geq 1}$ are valid for $\mu_{\:\!\rm r} = Q$,
the explicit generalization to $\mu_{\:\!\rm r} \neq Q\:\!$ to this order can 
be found in Eq.~(2.9) of Ref.~\cite{NV3}. 

The N$^{\:\!3}$LO physical kernels for the structure functions (\ref{Fns-def})
are not completely known at this point, as the four-loop splitting functions 
$P_{a,3}(x)$ contributing to Eq.~(\ref{Fevol}) have not been derived so far
beyond the small leading-$\nf$ contribution~\cite{Gracey:1994nn}.
Already the corresponding three-loop splitting functions $P_{a,2\,}$, however, 
have only a small impact at $\,x > 10^{-3}$, see Fig.~7 of Ref.~\cite{MVV3}. 
Moreover, the dependence of the non-singlet splitting function on $N$ and on 
the specific quark combination is such that a single four-loop moment of any of 
them sets the scale for the N$^{\:\!3}$LO contributions outside the small-$x$ 
region, cf.~Fig.~1 of Ref.~\cite{MVV3}.
Such a calculation has been presented in Ref.~\cite{Baikov:2006ai}, and the
fourth-order correction is indeed found to be small. Hence a rough estimate
of $P_{a,3}(x)$, for instance via an $N$-space Pad\'e estimate, is sufficient 
in Eq.~(\ref{Fevol}) for all practical non-singlet analyses.

The expressions for the transverse, longitudinal and asymmetric fragmentation 
functions
\beq
\label{FTns-def}
  {\cal F}_{T}^{} \:=\: F_{\:\! T,\rm ns}^{\, h}\:\: , \quad
  {\cal F}_{L}^{} \:=\: F_{\:\! L,\rm ns}^{\, h}\:\: , \quad
  {\cal F}_{\!A}^{} \:=\: F_{A}^{\, h} 
\eeq
in semi-inclusive $e^+e^-$ annihilation (see Ref.~\cite{PDG08} for a general
overview), $\, e^+e^- \,\ra\, \gamma \,/\, Z \,\ra\, h + X$, are completely 
analogous to those for the corresponding deep-inelastic structure functions. 
The scaling variable in Eq.~(\ref{Fns-cq}) now reads $x = 2pq/Q^2$ where $q$ 
with $q^{\:\!2} \equiv \Qs > 0$ is the momentum of the virtual gauge boson, and 
$p$ that of the identified hadron $h$.
The second-order non-singlet coefficient functions $c_{a,2}(x)$ for these 
cases have been calculated in Refs.~\cite{RvN96}, see also Ref.~\cite{MMV1} 
where we have derived the corresponding timelike splitting functions $P_{a,2}$ 
for the evolution of the non-singlet fragmentation densities $q_{a,\rm ns}$ of 
the hadron $h$. In these cases we know the three-loop coefficient functions 
$c_{a,3\,}$, beyond the leading large-$x$ terms of Refs.~\cite{Blumlein:2006pj,%
MV4}, only up some terms involving $\,\z2 = \pi^2/6$, cf.~the hadronic 
Higgs decay rate in Ref.~\cite{MV2}. These incomplete results have not been 
published. Their (complete) highest $\ln^{\,n} \!\x1$ terms will be presented
in the next section.

Finally we also consider the non-singlet quark-antiquark annihilation 
contribution to the total cross section for Drell-Yan lepton-pair production,
$pp/p\bar{p} \,\ra\, l^+l^-+X$, 
\beq
\label{DYns-def}
  {\cal F}_{\:\!\rm DY}^{} \:\: = \:\:
  \frac{1}{\sigma_0} \; \frac{d\:\! \sigma_{\rm ns}}{d\:\! \Qs} \;\; .
\eeq
In a rather schematic (but for our purpose sufficient) manner this quantity can
be written as
\beq
\label{DYns-cq}
  {\cal F}_{\:\!\rm DY}^{}(x,\Qs) 
  \:\: = \; \Big[ \, C_{\rm DY}(\Qs) \otimes q(\Qs) \otimes \bar{q}(\Qs) 
  \Big](x) \:\: . 
\eeq
Here $\Qs > 0$ denotes the squared invariant mass of the lepton pair, and the 
scaling variable is given by $x = \Qs/S$ where $S$ is the squared CMS energy
of the proton$\,$--$\,$(anti-)$\:\!$proton initial state. As in the case of 
deep-inelastic scattering, $q(x,\Qs)$ represents the initial-state (spacelike) 
quark densities of the proton which evolve with the splitting functions of 
Ref.~\cite{MV3}.
The non-singlet quark-antiquark coefficient function has a perturbative 
expansion analogous to Eq.~(\ref{Fns-cq}) above, with $\sigma_0^{}$ in Eq.~%
(\ref{DYns-def}) chosen such that also $c_{\rm DY,0}^{}(x)=\delta(1\!-\!x)$. 
The complete expressions for the NNLO contribution $c_{\rm DY,2}^{}(x)$ have 
been calculated in Refs.~\cite{Hamberg:1991np,Harlander:2002wh}. At N$^3$LO 
only the leading large-$x$ terms, $(1-x)^{-1} \ln^{\,n}(1-x)$ with 
$\,n = 0,\dots, 5$, are presently known from Ref.~\cite{MV1}, see also 
Ref.~\cite{Laenen:2005uz}.

The derivation of the physical evolution kernel for ${\cal F}_{\:\!\rm DY}^{}$
proceeds completely analogous to the paragraph of Eq.~(\ref{Fevol}), with the
non-singlet quark-quark splitting function occurring twice instead of once.
As we will show in the next section, this modification is irrelevant for the 
purpose of this article, the determination of subleading 
large-$x\,/\,$large-$N$ double logarithms in the higher-order coefficient 
functions for the quantities in Eqs.~(\ref{Fns-def}), (\ref{FTns-def}) and 
(\ref{DYns-def}).
%
%
\setcounter{equation}{0}
\section{Known large-{\boldmath $x$} logarithms at the second and third order}
\label{sec:largex1}
%
%
We next need to address the expansions in powers of $\ln \x1$ of the known 
non-singlet splitting functions \cite{FRS77,CFP80,MVV3,MMV1} and coefficient
functions \cite{ZvNcq2,Moch:1999eb,MVV2,%
MVV5,MVV6,MVV10,RvN96,Mitov:2006wy,Hamberg:1991np,Harlander:2002wh} in the
\MSb\ scheme. 
The spacelike splitting functions (\ref{qevol}) for all three types of 
non-singlet combinations of quark densities,
\beq
\label{qns-pmv}
  q_{\pm,{\rm ns}}^{\,(ik)} \; = \; q_i^{} \pm
  \bar{q}_i^{} - (q_k^{} \pm \bar{q}_k^{}) \;\; , \qquad
  q_{\rm v, ns}^{} \; = \; {\textstyle \sum_{\,r=1}^{\,\nf} } \,
  (q_r^{} - \bar{q}_r^{}) \:\: ,
\eeq
are given by
\beq
\label{Plargex}
  P_{a,l}(x) \; = \;
    \frac{A_{l\!+\!1}}{\x1_+} \: + \:  \widetilde{B}_{l\!+\!1}\: \delta \x1
    \: + \: \widetilde{C}_{l\!+\!1}\: \ln \x1 
    \: + \: {\cal O} \!\left( \x1^{\:\!k\, \geq 1} \ln^{\,l} \!\x1 \right) 
    \:\: .
\eeq
The last term indicates that the $(l\! +\! 1)$-loop splitting functions 
$P_{a,l}(x)$ receive contributions from terms no higher than $\ln^{\,l} \!\x1$, 
and that these contributions occur at all orders in $\x1$ from the first. 
It is interesting to note that the new colour structure $\dabc2$ entering 
the valence splitting function $P_{\rm v}$ at three loops contributes only 
non-leading terms $\x1^{\:\!k \geq 1} \ln \x1$, in striking contrast to its 
dominance in the small-$x$ limit \cite{MVV3}.
As indicated the first three terms in Eq.~(\ref{Plargex}) are the same for all
three splitting functions $P_a$ -- with the non-vanishing ($\,l>0$) coefficients
$\widetilde{C}_{l\!+\!1}$ being combinations of lower-order cusp anomalous 
dimensions $A_{n \leq l}$ -- and their functional form is independent on the 
perturbative order $l$. This independence is established to all orders in 
Ref.~\cite{Korchemsky:1989si} for the first two terms, and strongly suggested 
for the third term by the conjecture of Ref.~\cite{DMS05} and its third-order
verification in Refs.~\cite{MMV1,MV2}, see also Ref.~\cite{Basso:2006nk}.

Eq.~(\ref{Plargex}) also holds for the corresponding timelike splitting 
functions \cite{CFP80,MMV1} governing the evolution of the non-singlet 
fragmentation densities, with  the same large-$x$ coefficients except for a 
sign change of $\widetilde{C}$ relative to the spacelike case \cite{DMS05}. 
Hence it appears that none of the non-singlet splitting functions exhibits any 
large-$x$ double logarithms at any order of $\x1$.

The known coefficient functions for the deep-inelastic structure functions 
(\ref{Fns-def}) and the fragmentation functions (\ref{FTns-def}) 
-- with the obvious exception of ${\cal F}_L^{}$ -- 
receive the same highest double logarithmic contributions,
\beq
\label{strlargex}
  c_{a,l}(x) \:\: = \:\: \frac{1}{(l-1)!} \: (2\:\! \cf)^l \, p_{qq}(x) \,
  \ln^{\,2l-1} \!\x1 
  \: + \: {\cal O} \!\left( \x1^{\:\!k\, \geq -1} \ln^{\,2l-2} \!\x1 \right)
\eeq
with
\beq
\label{pqq0}
  p_{qq}(x) \:\: = \:\: \frac{2}{\x1_+}\: -\: 2 \:+\: \x1 
\eeq
and $C_F \,=\, (2\,N_c)^{-1\,} (N_c^{\,2}-1) \,=\, 4/3$ in QCD.
Eq.~(\ref{strlargex}) conforms to the general observation, going back to 
Ref.~\cite{KLS98}, that the coefficient of the highest $\x1^0$ logarithm is 
the negative of that of the highest +-distribution. Actually, this pattern 
also applies to the $C_F^{\,l-1}\,\{C_A,\,\nf\}$ $\ln^{\,2l-2} \!\x1$ and 
$C_F^{\,l-2} \{ \cas,\,C_A \nf,\, \nfs\}$ $\ln^{\,2l-3} \!\x1$ terms to $l = 3$,
see Refs.~\cite{MVV6,MVV10}. Analogous results, e.g., 
\beq
\label{DYlargex}
  c_{{\rm DY},l}^{}(x) \:\: = \:\: 
  \frac{1}{(l-1)!} \: (8\:\! \cf)^l \, p_{qq}(x) \, \ln^{\,2l-1} \!\x1
  \: + \: {\cal O} \!\left( \x1^{\:\!k\, \geq -1} \ln^{\,2l-2} \!\x1 \right)
\eeq
hold for the Drell-Yan cross section (\ref{DYns-def}).

For the convenience of the reader, we now proceed to provide the full 
$\ln^{\,n} \! (1-x)$ contributions to all known coefficient functions as far as
they are relevant for our main predictions in later sections. The 
coefficient functions (\ref{Fns-cq}) for ${\cal F}_1$ read
\bea
\label{c11ex}
 c_{1,1}^{}(x) 
  & = & 
      \ln \x1 \; 2\, \cf\, \pqq(x) 
  \nn \\[1mm]
  &   & \hspn \mbox{} 
      + \cf \* \left\{ \pqq(x)\, ( -3/2 - 2\, \H(0) ) 
          - \delta \x1\, ( 9 + 4\*\z2 ) + 3 + 3/2\: \x1 \right\} 
%
 \:\: , \\[3mm]
\label{c12ex}
 c_{1,2}^{}(x) 
  & = & 
        \left( \ln^{\,3} \!\x1 \; 4\, \cfs
          - \ln^{\,2} \!\x1 \, \cf\* \beta_0 \right)\*  \pqq(x)  
  \nn \\[1mm] 
  &   & \hspn \mbox{} 
      + \ln^{\,2} \!\x1 \, \left[ \cfs \* 
        \left\{ \pqq(x)\, ( - 9 - 14\, \H(0) ) 
                + 6 + 4\, \H(0) - \x1 \* ( 1 + 2\, \H(0) ) \right\} \right] 
  \nn \\[1mm]
  &   & \hspn \mbox{} 
      + \ln \x1 \; \left[ \cfs \*
        \left\{ \pqq(x)\, ( - 27/2 - 4\, \THh(1,0) + 24\, \Hh(0,0) 
                            + 12\, \H(0) - 16\,\z2 ) + 8\, \THh(1,0)
  \right. \right.  \nn \\
  &   & \hspp \mbox{} \left. \left.
        - 11 - 16\, \Hh(0,0) - 32\, \H(0) + \x1 \* ( - 53/2 - 4\*
          \THh(1,0) + 8\, \Hh(0,0) + 12\,\H(0) + 8\, \z2) \right\} \right.
  \nn \\[0.5mm]
  &   & \left. \mbox{} + \cf \* \beta_0 \*
       \left\{ \pqq(x)\, ( 29/6 + 4\, \H(0) ) - 3 + 5/2\: \x1 \right\} \right.
  \nn \\[1mm]
  &   & \left. \mbox{} + \cf \* \ca \* 
       \left\{ \pqq(x)\, ( 8/3 + 4\, \THh(1,0) + 4\, \Hh(0,0) - 4\, \z2 )
       + 2 + \x1 \* (14 - 4\, \z2) \right\} 
  \right. \nn \\[0.5mm]
  &   & \left. \mbox{} + \cf \* (\ca - 2\,\cf) \:
       \pqq(-x)\, ( 8\, \THh(-1,0) - 4\, \Hh(0,0) ) \right]
%
  \nn \\[1mm]
  &   & \hspn \mbox{}
      + {\cal O} \! \left( \:\!\ln^{\,0} \!\x1 \:\!\right) 
  \:\: , \\[2mm]
\label{c13ex}
 c_{1,3}^{}(x)
  & = & 
        \left( \ln^{\,5} \!\x1 \; 4\, \cft
      - \ln^{\,4} \!\x1 \; 10/3\: \cfs\,\beta_0 
      + \ln^{\,3} \!\x1 \; 2/3\: \cf \* \beta_0^2 \right) \* \pqq(x)
  \nn \\[1mm]
  &   & \hspn \mbox{}
      + \ln^{\,4} \!\x1 \, \left[ \cft \* 
        \left\{ \pqq(x)\, ( - 15 - 24\, \H(0) )
                + 6 + 8\, \H(0) - \x1 \* ( 5 + 4\, \H(0) ) \right\} \right] 
  \nn \\[1mm]
  &   & \hspn \mbox{}
      + \ln^{\,3} \x1 \; \left[ \cft \* 
        \left\{ \pqq(x)\, ( - 18 - 8\, \THh(1,0) + 296/3\: \Hh(0,0)
                            + 54\, \H(0) - 48\,\z2 ) + 32\, \THh(1,0)  
  \right. \right.  \nn \\
  &   & \hspp \mbox{} \left. \left.
        - 22 - 64\, \Hh(0,0) - 84\, \H(0) - \x1 \* ( 33 + 16\, 
          \THh(1,0) -32\, \Hh(0,0) - 54\,\H(0) - 16\, \z2) \right\} \right.
  \nn \\[0.5mm]
  &   & \left. \mbox{} + \cfs\, \beta_0 \* 
       \left\{ \pqq(x)\, ( 70/3 + 164/9\, \H(0) ) - 8 - 4\,\H(0) 
                         + \x1 \* ( 8 + 2\, \H(0) ) \right\} \right.
  \nn \\[1mm]
  &   & \left. \mbox{} + \cfs\, \ca \*
       \left\{ \pqq(x)\, ( 32/3 + 8\, \THh(1,0) +8\, \Hh(0,0) - 16\, \z2 )
       + 4 + \x1 \* ( 28 - 8\, \z2) \right\}
  \right. \nn \\[0.5mm]
  &   & \left. \mbox{} + \cfs \* (\ca - 2\,\cf) \:
       \pqq(-x)\, ( 16\, \THh(-1,0) - 8\, \Hh(0,0) ) \right]
%
  \nn \\[1mm]
  &   & \hspn \mbox{}
      + {\cal O} \! \left( \:\!\ln^{\,2} \!\x1 \:\!\right)
  \:\: .
\eea
Here and below we suppress the argument $x$ of the harmonic polylogarithms
\cite{Remiddi:1999ew} for brevity. Furthermore we use a slightly non-standard 
set of basis functions, i.e., Eqs.~(\ref{c12ex}) and (\ref{c13ex}) include 
\bea
  \THh(1,0)(x) \; & = & \; \Hh(1,0)(x) \: + \: \z2  
  \quad\:\: = \; - \ln x \, \ln \x1 - \mbox{Li}_2(x) + \z2
\:\: , \nn \\[1mm]
  \THh(-1,0)(x) & = & \Hh(-1,0)(x) + \z2/2 
  \; = \:\: \ln x \, \ln (1\!+\!x) + \mbox{Li}_2(-x) + \z2/2
  \nn 
\eea
besides $\H(0)(x) \,=\, \ln x$ and $\Hh(0,0)(x) \,=\, 1/2\: \ln^{\,2} \! x$.
All (modified) H-functions employed in our equations have a Taylor expansion
at $x=1$, starting at order $\x1$ or higher, with rational coefficients. Thus
also all terms with the Riemann $\zeta$-function can be read off directly from
our expansions. 

The corresponding results for the structure function ${\cal F}_2$ in 
(\ref{Fns-def}) are given by
\bea
\label{c21ex}
 c_{2,1}^{}(x)
  & = &
      c_{1,1}^{}(x) + 4\, x\:\cf
%
 \\[2mm]
\label{c22ex}
 c_{2,2}^{}(x)
  & = &
      c_{1,2}^{}(x) + \ln^{\,2} \!\x1 \: 8\, x\:\cfs 
  \nn \\[1mm]
  &   & \hspn \mbox{}
      + \ln \x1 \; \left[ \cfs\,
        \left\{ 8 - x \, ( 4 + 24\, \H(0) ) \right\}
        - 4\:\! x\:\cf \* \beta_0
      - \cf \* (\ca\!-\!2\,\cf) \: 16\:\! x\, ( 1 - \z2) \right]
  \nn \\[1mm]
  &   & \hspn \mbox{} 
      + \cfs \* \left\{ 12 - 8\,\H(0) - (130/3 + 8\,\THh(1,0) - 16\,\Hh(0,0) 
                       + 16\,\z2 )\: x \right\} 
  \nn \\[1mm]
  &   & \hspn \mbox{} 
      - \cf \* \beta_0\, \left\{ 4 - ( 50/3 + 8\,\H(0) )\: x \right\}
  \nn \\[1mm]
  &   & \hspn \mbox{} + \cf \* (\ca\!-\!2\,\cf) \* 
      \left\{ - 32/(5\,x^{\,2})\: ( \THh(-1,0) - \z2 /2)
       - 32/(5\,x)\: ( 1 - \H(0) )
  \right. \nn \\ & & \left. \mbox{} 
       + 8/5 - 32\,\THh(-1,0) - 16/5\: \H(0) + 16\,\z2
       + x \, ( 236/15 + 16\, [ 2\:\THhh(-1,-1,0) 
  \right. \nn \\[1mm] & & \left. \mbox{} 
            - \THhh(-1,0,0) + \THhh(1,0,0)
            - \THh(-1,0) + \Hh(0,0) ] + 104/5\,\H(0) - 8\,\z2 - 24\,\z3) 
  \right. \nn \\[1mm] & & \left. \mbox{} 
       - 48/5\: x^{\,2} \* ( 1 + \H(0) )
       + 48/5\: x^{\,3} \* ( \THh(-1,0) - \Hh(0,0) + \z2 /2) \right\}
%
  \:\: , \\[2mm]
\label{c23ex}
 c_{2,3}^{}(x)
  & = &
      c_{1,3}^{}(x) + \ln^{\,4} \!\x1 \: 8\, x\:\cft
  \nn \\[1mm]
  &   & \hspn \mbox{}
      + \ln^{\,3} \x1 \: \left[ \cft \*
        \left\{ 16 - x\, ( 8 + 48\, \H(0) ) \right\}
      - 32/3\,x\:\cfs \* \beta_0
      - \cfs \* (\ca\!-\!2\,\cf) \: 32\:\! x \, (1 - \z2) \right]
  \nn \\[1.5mm]
  &   & \hspn \mbox{}
      + \ln^{\,2} \x1 \; \Big[ \cft\,
      \left\{ 16 - 48\,\H(0) - (166/3 + 32\,\THh(1,0) - 160\,\Hh(0,0)
                       - 24\,\H(0) + 96\,\z2 )\: x \right\}
  \nn \\[1mm]
  &   & \mbox{}
      - \cfs\, \beta_0\, \left\{ 20 - ( 158/3 + 52\,\H(0) )\: x \right\}
      + \cf \* \beta_0^{\,2}\: 4\:\!x  
  \nn \\[1mm]
  &   & \mbox{} + \cfs \* (\ca\!-\!2\,\cf) \*
      \left\{ - 64/(5\,x^{\,2})\: ( \THh(-1,0) - \z2 /2)
      - 64/(5\,x)\: ( 1 - \H(0) )
  \right. \nn \\ & & \hspp \left. \mbox{}
      - 144/5 - 64\,\THh(-1,0) - 32/5\: \H(0) + 64\,\z2
      + x \, ( 872/15 + 32\, [ 2\,\THhh(-1,-1,0)
  \right. \nn \\[1mm] & & \hspp \left. \mbox{}
           - \THhh(-1,0,0) + \THhh(1,0,0)
           - \THh(-1,0) + \Hh(0,0) ] + 688/5\,\H(0) - 48\,\z2 
           - 96\,\z2\,\H(0) - 48\,\z3)
  \right. \nn \\[1mm] & & \hspp \left. \mbox{}
      - 96/5\: x^{\,2} \* ( 1 + \H(0) )
      + 96/5\: x^{\,3} \* ( \THh(-1,0) - \Hh(0,0) + \z2 /2) \right\} 
  \nn \\[-0.5mm]
  &   & \mbox{}
      + \cf \* (\ca\!-\!2\,\cf)\, \beta_0 \: 24\:\! x\, ( 1 - \z2) 
      + \cf \* (\ca\!-\!2\,\cf)^2 \: 32\:\! x\, (\z2 - \z3) \Big] \quad
%
  \nn \\[1mm]
  &   & \hspn \mbox{}
      + {\cal O} \! \left( \:\!\ln \x1 \:\!\right)
  \:\: .
\eea
The differences $c_{2,l}^{} - c_{1,l}^{}$ are, of course, the coefficient 
functions for the longitudinal structure function in DIS. Hence we have included
one more order in $\ln \x1$ in Eqs.~(\ref{c22ex}) and (\ref{c23ex}). This order
additionally includes a combination of three weight-three harmonic 
polylogarithms,
\bea
  \THhh(-1,-1,0)(x) & = & \Hhh(-1,-1,0)(x) \: + \: \H(-1)(x)\: \z2/2
  - \z3/8
 \:\: , \nn \\[1mm]
  \THhh(-1,0,0)(x) \; & = & \; \Hhh(-1,0,0)(x) - 3\,\z3/4
 \:\: , \nn \\[1mm]
  \THhh(1,0,0)(x) \;\; & = & \:\: \Hhh(1,0,0)(x) - \z3
 \nn
\eea
besides the unmodified $\Hhh(0,0,0)(x) \,=\, 1/6\: \ln^{\,3} \! x$.
The reader is referred to Ref.~\cite{Moch:1999eb} for expressions of these 
functions in terms of the standard polylogarithms Li$_2(x)$ and Li$_3(x)$.

To the same accuracy as Eqs.~(\ref{c11ex}) -- (\ref{c13ex}) for ${\cal F}_1$,
the coefficient functions for ${\cal F}_3$ can be written~as$\!$
\bea
\label{c31ex}
 c_{3,1}^{}(x)
  & = &
      c_{1,1}^{}(x) - 2\,\cf\, \x1
%
  \\[2mm]
\label{c32ex}
 c_{3,2}^{}(x)
  & = &
      c_{1,2}^{}(x) - \ln^{\,2} \x1 \; 4\,\cfs\, \x1
  \nn \\[1mm]
  &   & \hspn \mbox{}
      + \ln \x1 \; \left[ \cfs \* 
        \left\{ -32\, \H(0) + \x1 \* ( 18 + 28\, \H(0) - 16\,\z2 ) \right\}
  \right.  \nn \\[1.5mm]
  &   & \left. \mbox{}
      + 2\,\cf \* \beta_0\, \x1
      + \cf \* \ca \* \left\{ 16\, \H(0) - 8\,\x1 \* ( 1 + \H(0) - \z2) 
      \right\} 
  \right. \nn \\[0.5mm] 
  &   & \left. \mbox{} 
      - \cf \* (\ca - 2\,\cf) \,8\,\pqq(-x)\, (2\,\THh(-1,0) - \Hh(0,0) ) \right]
%
  \nn \\
  &   & \hspn \mbox{}
      + {\cal O} \! \left( \:\!\ln^{\,0} \!\x1 \:\!\right)
  \:\: , \\[2mm]
\label{c33ex}
 c_{3,3}^{}(x)
  & = &
      c_{1,3}^{}(x) - \ln^{\,4} \x1 \: 4\,\cft \,\x1
  \nn \\[1mm]
  &   & \hspn \mbox{}
      + \ln^{\,3} \x1 \; \left[ \cft \* 
        \left\{ -64\, \H(0) + \x1 \* ( 36 + 56\, \H(0) - 32\,\z2 ) \right\}
  \right.  \nn \\[1.5mm]
  &   & \left. \mbox{}
      + 16/3\:\cfs\, \beta_0\, \x1
      + \cfs\, \ca \* \left\{ 32\, \H(0) - 16\,\x1 \* ( 1 + \H(0) - \z2)
      \right\}
  \right. \nn \\[0.5mm]
  &   & \left. \mbox{}
      - \cfs\, (\ca - 2\,\cf) \,16\,\pqq(-x)\, ( 2\,\THh(-1,0) - \Hh(0,0) ) 
      \right]
%
  \nn \\
  &   & \hspn \mbox{}
      + {\cal O} \! \left( \:\!\ln^{\,2} \!\x1 \:\!\right)
 \:\: .
\eea
As mentioned below Eq.~(\ref{c13ex}), the H-functions in our expansions start
at order $\x1$ or higher at large $x$. Hence one can directly read off from 
Eqs.~(\ref{c31ex}) -- (\ref{c33ex}) that the coefficient functions for 
${\cal F}_1$ and ${\cal F}_3$ differ for $x\ra 1$ only in terms of order 
$\x1$. This fact was already noted in Ref.~\cite{MVV10}.

The third-order coefficient functions (\ref{c13ex}), (\ref{c23ex}) and 
(\ref{c33ex}) receive contributions from new flavour classes involving the 
higher group invariant $\dabc2$ \cite{MVV5,MVV6,MVV10}. The highest $\dabc2$
terms behave as $\x1 \,\ln \x1$ for ${\cal F}_1$ and ${\cal F}_2$, and as 
$\x1 \,\ln^{\,2} \! \x1$ for ${\cal F}_3$. Their leading contributions for the 
longitudinal structure function is of order $\x1^2 \,\ln \x1$. These terms will 
not be relevant on the level of our present analysis. The same holds for the
new three-loop functions $g_i(x)$ which also show only a single-logarithmic
behaviour for $x\ra 1$ \cite{MVV6,MVV10}.

The coefficient functions for the transverse fragmentation function ${\cal F}_T$
are related to those for ${\cal F}_1$ in DIS by suitably defined analytic
continuations. Hence we also present their $\ln \x1$ expansions relative to
the results for $c_{1,l}^{}(x)$ in Eqs.~(\ref{c11ex}) -- (\ref{c13ex}):
\bea
\label{cT1ex}
 c_{T,1}^{}(x)
  & = &
      c_{1,1}^{}(x) + \cf \* \left\{ 12\,\z2\, \delta \x1 
          + 6\,\pqq(x)\, \H(0) - 6  + 3\, \x1 \right\}
%
 \:\: , \\[2.5mm]
\label{cT2ex}
 c_{T,2}^{}(x)
  & = &
      c_{1,2}^{}(x) + \ln^{\,2} \x1 \: 
      \cfs\, \left\{ 24\,\pqq(x)\,\H(0) - 12 + 6\,\x1 \right\} 
  \nn \\[1mm]
  &   & \hspn \mbox{}
      + \ln \x1 \; \left[ \cfs \* 
        \left\{ \pqq(x)\, ( 8\,\THh(1,0) - 28\,\Hh(0,0) - 18\,\H(0) +24\,\z2 )
               \right. \right.
  \nn \\[1mm]
  &   & \left. \left. \hspp \mbox{}
        + 22 + 40\,\Hh(0,0) + 12\,\H(0) 
        - \x1 \* ( 11 + 20\,\Hh(0,0) + 10\,\H(0)  ) 
      \right\}
  \right.  \nn \\[1mm]
  &   & \left. + \cf \* \beta_0 \* 
        \left\{ -6\,\pqq(x)\,\H(0) + 6 - 3\,\x1 \right\} 
                \right.
  \nn \\[1mm]
  &   & \left. + \cf \* \ca \* 
        \left\{ \pqq(x)\, ( - 8\,\THh(1,0) - 4\,\Hh(0,0) ) - 4 + 2\,\x1 
      \right\} \right]
%
  \nn \\[1mm]
  &   & \hspn \mbox{}
      + {\cal O} \! \left( \:\!\ln^{\,0} \!\x1 \:\!\right)
  \:\: , \\[2mm]
\label{cT3ex}
 c_{T,3}^{}(x)
  & = &
      c_{1,3}^{}(x) + \ln^{\,4} \x1 \:
      \cft\, \left\{ 36\,\pqq(x)\,\H(0) - 12 + 6\,\x1 \right\}
  \nn \\[0.5mm]
  &   & \hspn \mbox{}
      + \ln^{\,3} \x1 \; \left[ \cft \* 
        \left\{ \pqq(x)\, ( 16\,\THh(1,0) - 104\,\Hh(0,0) - 72\,\H(0) 
                + 48\,\z2 ) \right. \right.
  \nn \\[1mm]
  &   & \left. \left. \hspp \mbox{}
        + 44 + 128\,\Hh(0,0) + 24\,\H(0) 
        - \x1 \* ( 22 + 64\,\Hh(0,0) + 44\,\H(0)  )
      \right\}
  \right.  \nn \\[2mm]
  &   & \left. + \cfs\, \beta_0 \* 
        \left\{ -232/9\:\pqq(x)\,\H(0) + 16 - 8\,\x1 \right\}
                \right.
  \nn \\[1mm]
  &   & \left. + \cfs\, \ca \* 
        \left\{ \pqq(x)\, ( - 16\,\THh(1,0) - 8\,\Hh(0,0) ) - 8 + 4\,\x1 
      \right\} \right]
%
 \nn \\[1mm]
  &   & \hspn \mbox{}
      + {\cal O} \! \left( \:\!\ln^{\,2} \!\x1 \:\!\right)
  \:\: .
\eea
As for the spacelike case of Eqs.~(\ref{c21ex}) -- (\ref{c23ex}), also the
coefficient functions for the timelike longitudinal structure function 
${\cal F}_L$ in Eq.~(\ref{FTns-def}) will be needed to one more order in 
$\ln \x1$ below. Their corresponding expansions are given by
\bea
\label{cL1tex}
 c_{L,1}^{}(x)
  & = &
      2\, \cf
%
  \:\: ,\\[2mm]
\label{cL2tex}
 c_{L,2}^{}(x)
  & = &
      \ln^{\,2} \x1 \; 4\, \cfs
  \nn \\[1mm]
  &   & \hspn \mbox{}
      + \ln \x1 \: \left[ \cfs \* 
        \left\{ -2 + 4\,\H(0) + 4\:\! x \right\} - 2\,\cf \* \beta_0 
        - 8\,\cf \* (\ca\!-\!2\,\cf) \* ( 1 - \z2) \right]
   \nn \\[1mm]
  &   & \hspn \mbox{}
      + \cfs \* \left\{ - 41/3 - 12\,\THh(1,0) - 12\,\Hh(0,0) + 2\,\H(0)
                       + 16\,\z2 - x\, (2 - 8\,\H(0) ) \right\}
  \nn \\[1mm]
  &   & \hspn \mbox{}
      + \cf \* \beta_0\, \left\{ 25/3 - 2\,\H(0) - 2\,x \right\}
  \nn \\[1mm]
  &   & \hspn \mbox{} + \cf \* (\ca\!-\!2\,\cf) \* 
      \left\{ - 24/(5\,x^{\,2})\: ( \THh(-1,0) - \z2 /2)
       - 24/(5\,x)\: ( 1 - \H(0) ) + 118/15
  \right. \nn \\ & & \left. \mbox{}
       - 8 \* [ 2\,\THhh(-1,-1,0) - \THhh(-1,0,0) - 2\, \THhh(0,-1,0)
       - \THhh(1,0,0) - \THh(-1,0) ] - 12/5\:\H(0)
  \right. \nn \\[1mm] & & \left. \mbox{}
       - 4\,\z2 - 8\,\z2\,\H(0) - 12\,\z3 
       + x\, ( 4/5 + 16\,\THh(-1,0) - 16\,\Hh(0,0) + 8/5\,\H(0) + 8\,\z2 )
  \right. \nn \\[1mm] & & \left. \mbox{}
       - 16/5\: x^{\,2} \* ( 1 + \H(0) )
       + 16/5\: x^{\,3} \* ( \THh(-1,0) - \Hh(0,0) + \z2 /2) \right\}
%
  \:\: , \\[2mm]
\label{cL3tex}
 c_{L,3}^{}(x)
  & = &
      \ln^{\,4} \x1 \: 4\,\cft
  \nn \\[1mm]
  &   & \hspn \mbox{}
      + \ln^{\,3} \x1 \; \left[ \cft \*
        \left\{ - 4 + 8\,\H(0) + 8\:\! x \right\}
      - 16/3\:\cfs\, \beta_0
      - 16\,\cfs\, (\ca\!-\!2\,\cf) \* ( 1 - \z2 ) \right]
  \nn \\[1.5mm]
  &   & \hspn \mbox{}
      + \ln^{\,2} \x1 \; \Big[ \cft\,
      \left\{ - 35/3 -16\,\THh(1,0) - 16\,\Hh(0,0) 
              - x\, ( 8 - 16\,\H(0) ) \right\}
  \nn \\[1mm]
  &   & \mbox{}
      + \cfs\, \beta_0\, \left\{ 79/3 - 10\,\H(0) - 10\:\! x \right\}
  \nn \\[1mm]
  &   & \mbox{} + \cfs \* (\ca\!-\!2\,\cf) \* 
      \left\{ - 48/(5\,x^{\,2})\: ( \THh(-1,0) - \z2 /2)
       - 48/(5\,x)\: ( 1 - \H(0) ) + 436/15
  \right. \nn \\ & & \hspp \left. \mbox{}
       - 16 \* [ 2\,\THhh(-1,-1,0) - \THhh(-1,0,0) - 2\, \THhh(0,-1,0)
       - \THhh(1,0,0) - \THh(-1,0) ] - 104/5\: \H(0) 
  \right. \nn \\[1mm] & & \hspp \left. \mbox{}
       - 24\,\z2 - 24\,\z3 + x \, ( -72/5 + 32\, \THh(-1,0) - 32\,\Hh(0,0) 
       + 16/5\,\H(0) + 32\,\z2 ) 
  \right. \nn \\[1mm] & & \hspp \left. \mbox{}
       - 32/5\: x^{\,2} \* ( 1 + \H(0) )
       + 32/5\: x^{\,3} \* ( \THh(-1,0) - \Hh(0,0) + \z2 /2) \right\}
  \nn \\[-0.5mm]
  &   & \mbox{}
       + 2\, \cf \* \beta_0^{\,2}
       + \cf \* (\ca\!-\!2\,\cf)\, \beta_0 \: 12\, ( 1 - \z2)
       + \cf \* (\ca\!-\!2\,\cf)^2 \: 16\, (\z2 - \z3) \Big] \quad
%
  \nn \\[1mm]
  &   & \hspn \mbox{}
      + {\cal O} \! \left( \:\!\ln \x1 \:\!\right)
  \:\: .
\eea
The last two equations include one more modified harmonic polylogarithm,
$$
  \THhh(0,-1,0)(x) \;\; = \;\; 
  \Hhh(0,-1,0)(x) + \H(0)(x)\: \z2/2 + 3\,\z3/2 
%
\:\: .
$$

The final observable in Eq.~(\ref{FTns-def}), the asymmetric fragmentation
function ${\cal F}_A$, is analogous to the structure function ${\cal F}_3$ 
in DIS. The $\as$-expansion (\ref{Fns-cq}) of its coefficient function reads
\bea
\label{cA1ex}
 c_{A,1}^{}(x)
  & = &
      c_{T,1}^{}(x) - 2\,\cf\, \x1
%
  \:\: , \\[2mm]
\label{cA2ex}
 c_{A,2}^{}(x)
  & = &
      c_{T,2}^{}(x) - \ln^{\,2} \x1 \; 4\,\cfs\, \x1
  \nn \\[1mm]
  &   & \hspn \mbox{}
      + \ln \x1 \; \left[ \cfs \* 
        \left\{ -32\, \H(0) + \x1 \* ( 18 + 12\, \H(0) - 16\,\z2 ) \right\}
  \right.  \nn \\[1.5mm]
  &   & \left. \mbox{}
      + 2\,\cf \* \beta_0\, \x1
      + \cf \* \ca \* \left\{ 16\, \H(0) - 8\,\x1 \* ( 1 + \H(0) - \z2)
      \right\}
  \right. \nn \\[0.5mm]
  &   & \left. \mbox{}
      - \cf \* (\ca - 2\,\cf) \,8\,\pqq(-x)\, (2\,\THh(-1,0) - \Hh(0,0) ) \right]
%
  \nn \\
  &   & \hspn \mbox{}
      + {\cal O} \! \left( \:\!\ln^{\,0} \!\x1 \:\!\right)
  \:\: , \\[2mm]
\label{cA3ex}
 c_{A,3}^{}(x)
  & = &
      c_{T,3}^{}(x) - \ln^{\,4} \x1 \: 4\,\cft \,\x1 
  \nn \\[1mm]
  &   & \hspn \mbox{}
      + \ln^{\,3} \x1 \; \left[ \cft \* 
        \left\{ -64\, \H(0) + \x1 \* ( 36 + 24\, \H(0) - 32\,\z2 ) \right\}
  \right.  \nn \\[1.5mm]
  &   & \left. \mbox{}
      + 16/3\: \cfs\, \beta_0\, \x1
      + \cfs \* \ca \* \left\{ 32\, \H(0) - 16\,\x1 \* ( 1 + \H(0) - \z2)
      \right\}
  \right. \nn \\[0.5mm]
  &   & \left. \mbox{}
      - \cfs\, (\ca - 2\,\cf) \,16\,\pqq(-x)\, (2\,\THh(-1,0) - \Hh(0,0) ) 
      \right]
%
  \nn \\
  &   & \hspn \mbox{}
      + {\cal O} \! \left( \:\!\ln^{\,2} \!\x1 \:\!\right)
  \:\: .
\eea
The relations (\ref{cT3ex}), (\ref{cL3tex}) and (\ref{cA3ex}) for the 
third-order timelike coefficient functions have not been presented before. 
These results have been obtained by extending the analytic continuations of 
Ref.~\cite{MMV1} to terms of order $\as^{\,3}\, \ep^{\,0}$ in dimensional 
regularization. At the end of this section we will present sufficient evidence 
that, for these $\ln \x1$ contributions, this analytic continuation does not 
suffer from the $\pi^2$-problem mentioned below Eq.~(\ref{FTns-def}).

Finally the known coefficient functions for the Drell-Yan cross section
(\ref{DYns-def}) are given~by
\bea
\label{cD1ex}
 c_{\rm DY,1}^{}(x) 
  & = & 
      \ln \x1 \; 8\, \cf\, \pqq(x) 
      + \cf \* \left\{ - 4\,\pqq(x)\,\H(0)  
          - \delta \x1\, ( 16 - 8\,\z2 ) \right\} 
%
 \:\: , \\[3mm]
\label{cD2ex}
 c_{\rm DY,2}^{}(x) 
  & = & 
        \left( \ln^{\,3} \x1 \; 64\, \cfs
      - 8\, \ln^{\,2} \x1 \, \cf \* \beta_0 \right) \* \pqq(x)  
  \nn \\[1mm] 
  &   & \hspn \mbox{} 
      + \ln^{\,2} \x1 \, \left[ \cfs \* 
        \left\{ - 124\,\pqq(x)\, \H(0) 
                + 64\, \H(0) - \x1 \* ( 64 + 32\, \H(0) ) \right\} \right] 
  \nn \\[1mm]
  &   & \hspn \mbox{} 
      + \ln \x1 \; \left[ \cfs \* 
        \left\{ \pqq(x)\, ( - 128  - 8\, \THh(1,0) + 112\, \Hh(0,0) - 24\,\H(0)
                            - 64\,\z2 ) - 4 
  \right. \right.  \nn \\
  &   & \hspp \mbox{} \left. \left.
        + 96\, [ \THh(1,0) - \Hh(0,0) ] - 160\, \H(0) + \x1 \* ( 4 - 48\,
          \THh(1,0) + 48\, \Hh(0,0) + 168\,\H(0) ) \right\} \right.
  \nn \\[0.5mm]
  &   & \left. \mbox{} + \cf \* \beta_0 \* 
       \left\{ \pqq(x)\, ( 40/3 + 16\, \H(0) ) + 16\, \x1 \right\} \right.
  \nn \\[1mm]
  &   & \left. \mbox{} + \cf \* \ca \* 
       \left\{ \pqq(x)\, ( 32/3 + 8\, \THh(1,0) + 16\, \Hh(0,0) - 16\, \z2 )
       + 4 + 16\, \THh(1,0) 
  \right. \right.  \nn \\
  &   & \hspp \mbox{} \left. \left.
       + 32\, \H(0) + \x1 \* (44 - 8\,\THh(1,0) - 16\, \H(0) ) 
       \right\} \right]
%
  \nn \\[1mm]
  &   & \hspn \mbox{}
      + {\cal O} \! \left( \:\!\ln^{\,0} \!\x1 \:\!\right) 
  \:\: .
\eea
At the third order only the +-distributions contributions 
$[ \:\!\x1^{-1}\, \ln^{\,n} \!\x1 \:\! ]_+^{}$, $ n = 0,\,\ldots,\,5$ 
are known so far, see Ref.~\cite{MV1}.

Eqs.~(\ref{c11ex}) --  (\ref{cA3ex}) can be employed to derive the $\ln \x1$
expansion of the physical evolution kernels (\ref{Fevol}) for the 
deep-inelastic structure functions (\ref{Fns-def}) and the $e^+e^-$ 
fragmentation functions ${\cal F}_T$, ${\cal F}_T+{\cal F}_L$ and ${\cal F}_A$ 
of Eq.~(\ref{FTns-def}). Recalling that $K_{a,n}$ denotes the 
N$^{\:\!n\:\!}$LO kernel for ${\cal F}_a$, one finds 
\bea
 K_{a,0}^{}(x)
  & = &
      2\,\cf \,\pqq(x) + 3\,\cf\, \delta \x1 
%
\:\: ,
  \nn \\[2mm]
 K_{a,1}^{}(x)
  & = &
  \ln \x1 \,\pqq(x) \* \left[
      -2\,\cf \* \beta_0 \:\mp\: 8\,\cfs\, \H(0) \right]
%
      \;+\; {\cal O} \! \left( \:\!\ln^{\,0} \!\x1 \:\!\right) \:\: ,
  \nn \\[2mm]
 K_{a,2}^{}(x)
  & = &
      \ln^{\,2} \x1 \,\pqq(x) \* \left[ 
      2\,\cf \* \beta_0^{\,2} \:\pm\: 12\,\cfs\,\beta_0\, \H(0) 
      + 32\,\cft\: \Hh(0,0) \right]
%
     \;+\; {\cal O} \! \left( \:\!\ln \x1 \:\!\right) \:\: ,
  \nn \\[2mm]
\label{Ka3}
 K_{a,3}^{}(x)
  & = &
      \ln^{\,3} \x1 \,\pqq(x) \* \left[
      - 2\,\cf \* \beta_0^{\,3} \:\mp\: 44/3\: \cfs\,\beta_0^{\,2}\: \H(0) 
      - 64\,\cft\,\beta_0\: \Hh(0,0) + \xi_{P_3}^{}\, \cff\: \Hhh(0,0,0) 
      \right]
%
  \nn \\[1mm]
  &   & \hspn \mbox{}
      +\; {\cal O} \! \left( \:\!\ln^{\,2} \!\x1 \:\!\right)
  \:\: .
\eea
From the NLO result $K_{a,1}^{}$ we have only written down the leading 
$\ln \x1$ terms. These contributions are the same for all six structure 
functions up to a sign change of the $\H(0)$ terms between the DIS quantities 
(upper sign) and the fragmentation functions (lower sign). The non-$\beta_0$
terms in Eqs.~(\ref{Ka3}) are the contributions of the \MSb\ splitting 
functions (\ref{Plargex}), consequently the fourth-order coefficient 
$\xi_{P_3}$ is unknown at this point, but irrelevant for our further 
considerations.

The corresponding results for the Drell-Yan cross section (\ref{DYns-def}) are 
given by
\bea
 K_{\rm DY,0}^{}(x)
  & = &
      4\,\cf \,\pqq(x) + 6\,\cf\, \delta \x1 
%
      \:\: ,
  \nn \\[2mm]
\label{KDY2}
 K_{\rm DY,1}^{}(x)
  & = & 
  \ln \x1 \,\pqq(x) \* \left[
      -8\,\cf \* \beta_0 - 16\,\cfs\, \H(0) \right]
%
      \;+\; {\cal O} \! \left( \:\!\ln^{\,0} \!\x1 \:\!\right) \:\: ,
  \\[2mm]
 K_{\rm DY,2}^{}(x)
  & = &
      \ln^{\,2} \x1 \,\pqq(x)\* \left[ 
      16\,\cf \* \beta_0^{\,2} + 56\,\cfs\,\beta_0\, \H(0) 
      + 64\,\cft\: \Hh(0,0) \right]
%
      \;+\; {\cal O} \! \left( \:\!\ln \x1 \:\!\right) 
  \nn \:\: .
\eea

Eqs.~(\ref{Ka3}) and (\ref{KDY2}) represent our crucial observation: 
the physical kernels for all seven non-singlet observables display an only 
single-logarithmic large-$x$ enhancement, at all powers of $\x1$, to all orders 
in $\as$ for which the corresponding coefficient functions are known. We 
consider it extremely unlikely that this pattern is accidental, and hence
conjecture a single-logarithmic behaviour of these physical at (all) higher 
orders in $\as$, with the leading contribution showing the same independence on
the specific structure function as in Eqs.~(\ref{Ka3}). 
In support of this conjecture we note that for the +-distribution parts of 
$K_{a,n}$ (including the Drell-Yan case), recall Eq.~(\ref{pqq0}), this 
single-logarithmic enhancement is established by the soft-gluon exponentiation 
as explained in the next section. Furthermore the all-order leading-$\nf$
results of Ref.~\cite{Mankiewicz:1997gz} prove the all-order generalization of 
Eqs.~(\ref{Ka3}) for the $C_F \beta_0^{\,n\,}$ contributions to the DIS kernels
$K_{1,n}$ and $K_{2,n}$.

The single-logarithmic enhancement of the physical kernels directly leads to
predictions for the highest $\ln \x1$ terms of the higher-order coefficient
functions. Considering, for example, the third (N$^3$LO) line of Eqs.~(\ref
{ctilde}), one notes that the convolutions of $c_{a,1}(x)$ and $c_{a,2}(x)$ 
lead to terms up to $\ln^{\,5}\!\x1$. The vanishing of terms higher than 
$\ln^{\,3} \x1$ thus fixes the $\ln^{\,4}\!\x1$ and $\ln^{\,5}\!\x1$
terms of $c_{a,3}(x)$. In fact, exactly this reasoning, together with the
absence of any $\z2$ terms in Eqs.~(\ref{Ka3}), provides the additional
confirmation of the correctness of Eqs.~(\ref{cT3ex}), (\ref{cL3tex}) and 
(\ref{cA3ex}) mentioned below the latter equation.
%
%
\setcounter{equation}{0}
\section{Soft-gluon exponentiation of the leading contributions}
\label{sec:softglue}
%
%
The leading (+-distribution) large-$x$ terms of the above coefficient functions
can be expressed to all orders in $\as$ in terms of the soft-gluon 
exponentiation \cite{SoftGlue}. Switching to the Mellin moments defined in 
Eq.~(\ref{Ndef2}), these contributions to Eq.~(\ref{Fns-cq}) can be written as
\beq
\label{CN-res} 
  C(N) \:\: = \:\: g_0^{}(\ar) \,\exp \Big\{\ln N\, g_1^{}(\lam) 
  + g_2^{}(\lam) + \ar\, g_3^{}(\lam) + {\cal O}(\ar^{\,2\,} f(\lam)) \Big\}
\eeq
up to terms which vanish for $N \!\rightarrow \!\infty$. Here we have
used the standard abbreviation
\beq
\label{lam-def} 
  \lambda \:\: =\:\: \ar \beta_0 L 
  \:\: \equiv \:\: \frac{\as}{4\pi} \:\beta_0 L \quad \mbox{ with } \quad
  L \:\: \equiv \:\: \ln N \:\: ,
\eeq
and we have again put $\mu_{\:\!\rm r\,} = \mu_{\:\!\rm f\,}^{} = Q$. By virtue
of the first line of Eq.~(\ref{Fevol}) -- the logarithmic derivative in
$N$-space -- Eq.~(\ref{CN-res}) leads to the following expression for the 
resummed kernels up to next-to-next-to-leading logarithmic (NNLL) accuracy 
\cite{NV3}:
\bea
\label{KN-res} 
  K_a(N) &\, =\, & \mbox{} 
  - \eta_a \left( A_1\,\ar + A_2\,\ar^{\,2} + A_3\,\ar^{\,3} \right) \, \ln N 
  \: - \: \bigg( 1+ \frac{\beta_1}{\beta_0}\: \ar + \frac{\beta_2}{\beta_0}\:
  \ar^{\,2} \bigg)\, \lambda^2 \,\frac{dg_{a,1}}{d\lambda}
  \nonumber \\ & & \mbox{}
  - \Big( \ar \beta_0 + \ar^{\,2} \,\beta_1 \Big)\, \lambda
  \frac{dg_{a,2}}{d\lambda} \: - \: \ar^{\,2} \,\beta_0\,
  \frac{d}{d\lambda} \Big( \lambda g_{a,3}(\lambda)\Big)
  \: + \: {\cal O}(a_s^3 (f(\lam)) 
\eea
with $\,\eta_a = 2\,$ for $a = $ DY and $\,\eta_a = 1\,$ otherwise, cf.~the
last paragraph of Section 2. Thus the leading logarithmic (LL), next-to-leading
logarithmic (NLL) and NNLL large-$N$ contributions to the physical kernels are 
of the form $(a_s \ln N)^n$, $a_s (a_s \ln N)^n$ and $a_s^2 (a_s \ln N)^n$, 
respectively. Recalling 
$$
  f(N) \; = \; \frac{(-1)^n}{n}\: \ln^{\,n} N \:+\: {\cal O}(\ln^{\,n-2} N) 
  \quad \mbox{ for } \quad
  f(x) \; = \; \left[ \frac{\ln^{\,n-1} \! \x1}{1-x} \right]_+ \:\: ,
$$
one notes that Eq.~(\ref{KN-res}) implies that the single-logarithmic 
enhancement (\ref{Ka3}) and (\ref{KDY2}) holds to all order in $\as$ for the
+-distribution contributions.

In the next two sections we will provide analogous all-order results for the
subleading $\,N^{-1} \ln^{\,n} N$ contributions to the coefficient functions.
These results, however, will be restricted to a tower-expanded NLL accuracy, 
see Ref.~\cite{av99}. The exponents analogous to Eq.~(\ref{CN-res}) will be
given relative to the LL and  NLL functions entering the soft-gluon
exponent. For the deep-inelastic structure functions ${\cal F}_{1,\,2,\,3}$ and
the fragmentation functions ${\cal F}_{T\!+L,\,T,\,A}$ these functions read
\bea
\label{g1n}
  g_{a,1}^{}(\ar L) &\! = & \frac{A_1}{\beta_0 \lam}\,
    \Big[ \,\lam + (1-\lam) \ln(1-\lam) \Big]
    \;\;\equiv\;\; \sum_{k=1} g_{1k}^{}(\ar L)^k
    \nn \\ &\! = & 
    \sum_{k=1}^{\infty}\: \frac{A_1 \beta_0^{\,k-1}}{k(k+1)}
    \: (\ar L)^k \:\: , \\[3mm]
\label{g2n}
  g_{a,2}^{}(\ar L) &\! =\! & - \frac{\GE\,A_1 - B_1}{\beta_0}\, \ln(1-\lam)
    - \frac{A_2}{\beta_0^2}\, \Big[ \lam + \ln(1-\lam)\Big]
  \nonumber \\[0.5mm]
  & & \mbox{}+ \frac{A_1 \beta_1}{\beta_0^3}\,
    \Big[ \lam + \ln (1-\lam) + \frac{1}{2}\ln^2(1-\lam) \Big]
    \:\:\equiv\:\: \sum_{k=1} g_{2k}^{}(\ar L)^k
  \nn \\[0.5mm]
  &\! =\! & \sum_{k=1}^{\infty} \left\{ \frac{\GE\,A_1 - B_1}{\beta_0}
    + \theta_{k2} \!\left( \frac{A_2}{\beta_0^{2}}
    + \frac{A_1 \beta_1^{}}{\beta_0^{3}} \Big[ S_1(k-1)-1 \Big]
      \right) \right\} \frac{\beta_0^{\,k}}{k} \, (\ar L)^k 
\eea
with $\,\theta_{kj} = 1\,$ for $\,k\geq j\,$ and $\,\theta_{kj} = 0\,$ else, 
and $\,S_1(k) = \sum_{j=1}^{k} 1/j$ \cite{av99}. Here $A_1$ and $A_2$ are the
one- and two-loop +-distribution coefficients in Eq.~(\ref{Plargex}), given by 
\cite{Kodaira:1982nh}
\beq
\label{cusp12} 
   A_1 \:\: = \:\: 4\, C_F
   \;\; , \quad
   A_2 \:\: = \:\: 8\, C_F\, K
       \:\: = \:\: 8\, C_F  \Big[ \Big(\,\frac{67}{18} - \z2 \Big)\, C_A
                   - \frac{5}{9}\: \nf \Big] \:\: .
\eeq
Note that, besides these expansion coefficients and those of the beta function
(\ref{asrun}), only one additional coefficient,
\beq
  B_1 \:\: = \:\: - 3\, C_F
\eeq
enters the function $g_{a,2}$ in Eq.~(\ref{g2n}) \cite{SoftGlue}. 
This pattern does persist at
higher orders of the exponentiation \cite{MVV2,FR05GR05}. 
Consequently the functions $g_{n>1}$ are completely fixed by the first term 
of their respective expansion in $\as$, if the cusp anomalous dimension and 
the beta function are known to a sufficient accuracy. To a large extent the
predictive power of the soft-gluon exponentiation rests on this fact: the
calculation of the N$^{\,n}$LO coefficient function is sufficient to also 
determine the N$^{\,n}$LL resummation function $g_{n+1}$, and thus two 
additional all-order towers of logarithms, see, e.g., Ref.~\cite{MVV7}. 
As we will see below, however, this situation does not directly generalize to 
the non-leading large-$x/\,$large-$N$ terms addressed in the next two sections. 

The LL and NLL resummation exponents for the Drell-Yan cross section 
(\ref{DYns-def}) are related to Eqs.~(\ref{g1n}) and (\ref{g2n}) as follows: 
\bea
    g_{\rm DY, 1}^{}(\lambda) & = &  \! 2\, g_{\rm DIS, 1}^{} (2\lambda) \;\; ,
 \nn \\[1mm]
    g_{\rm DY, 2}^{}(\lambda) & = &  \;\; g_{\rm DIS, 2}^{} (2\lambda)
 \quad \mbox { with } \quad B_1 \:\ra\: 0 \;\mbox { and }\; \GE \:\ra\: 2\,\GE
 \:\: .
\eea
Here, as above, $\GE\,\simeq\, 0.577216$ denotes the Euler-Mascheroni constant.
The absence of any NLL resummation coefficient additional to $A_1$ and 
$\beta_1$ is a low-order `accident', non-vanishing coefficients 
$D_{\:\!\rm DY, n}$ occur at NNLL \cite{av00} and all higher orders.

For later convenience we finally recall the leading contributions $g_{31}^{}$,
defined as for $g_1^{}$ and $g_2^{}$ in Eqs.~(\ref{g1n}) and (\ref{g2n}), to 
the NNLL resummation function in Eq.~(\ref{CN-res}). The universal coefficient 
for DIS and $e^+e^-$ annihilation, first extracted in Ref.~\cite{av99} from the
NNLO result of Refs.~\cite{ZvNcq2}, reads  
\bea
\label{g31}
  g_{a,31}^{} &\! =\! &
    \left( \frac{3155}{54} - \frac{22}{3}\: \z2 - 40\, \z3 
    - 8\,\z2\, \GE + \frac{22}{3}\: \GE^{\,2} 
    + \frac{367}{9}\: \GE\! \right) C_F C_A
  \nn \\[1.5mm] & & \hspn \mbox{} 
   \;+\; \left( \frac{3}{2} - 12\, \z2 + 24\, \z3\! \right) \cfs
   \;-\; \left( \frac{247}{27} - \frac{4}{3}\: \z2 + \frac{4}{3}\,
    \GE^{\,2} + \frac{58}{9}\: \GE\! \right) C_F \nf\:\: .
\eea
The corresponding result for the Drell-Yan case \cite{av00,Hamberg:1991np}
is given by
\bea
\label{g31DY}
  g_{\rm DY,31}^{} &\! =\! &
    \left( \frac{1616}{27} - 56\,\z3
     - 32\,\z2 \,\GE + \frac{176}{3}\: \GEs
     + \frac{1072}{9}\: \GE\! \right) C_F C_A \qquad
  \nn \\[1.5mm] & & \hspn \mbox{} 
    \;-\; \left( \frac{224}{27} + \frac{32}{3}\: \GE^{\,2} 
     + \frac{160}{9}\: \GE\! \right) C_F \nf\:\: .
\eea
Having collected all relevant fixed-order and soft-gluon resummation 
information, we can now turn to our new higher-order predictions.
%
%
\setcounter{equation}{0}
\section{Non-leading large-{\boldmath $N /$\,large-$x$} terms in structure
 functions}
\label{sec:largex2}
%
%
Keeping only the leading and subleading contributions, the large-$N$
behaviour of the coefficient functions (\ref{Fns-cq}) for the structure 
functions ${\cal F}_a$ in Eq.~(\ref{Fns-def}) can be written as
\beq 
\label{dcoeffs}
  c_{a,n}(N) \;\: = \;\: \sum_{k=0}^{2n} \,c_{nk} \: L^k 
  \; + \; {1 \over N}\, \sum_{k=0}^{2n-1} d^{\,(n)}_{a,\,k}\; L^k
  \: + \: O\Big( \, {1 \over N^{\,2}}\: L^{2n-1\,} \Big)
\eeq
with, as in the previous section, $L \,\equiv\, \ln N$. At the present accuracy 
the leading soft-gluon coefficients do not depend on the structure function, 
thus we have written $c_{nk\,}$ instead of $c^{\,(n)}_{a,\,k\,}$. 
$C_1$ and $C_3$ are identical at the level of Eq.~(\ref{dcoeffs}) as discussed
below Eq.~(\ref{c33ex}) -- recall that $N^{-2} \ln^{\,a} N$ corresponds to 
$\x1\, \ln^{\,a} \!\x1$.
Note that the second sum extends to $2n-1$, i.e., higher by one than the 
corresponding expansion for $\FL$ analysed in Ref.~\cite{MV3}. Thus the highest
coefficients $d^{\,(n)}_{a,\,2n-1}$ at each order $n$ are identical also for 
$C_1$ and $C_2$. Recall that also the leading logarithms of the physical 
evolution kernels (\ref{Ka3}) to N$^{\,3}$LO are the same for $a = 1,\,2,\,3$.
Their $1/N$ contributions are
\bea
\label{KaLLN}
  K_{\,a,1} \Big|_{\,N^{-1}L}\; & \!= \! & \mbox{}
     - 2\,\beta_0\, \cf \:-\: 16\,\cfs
\:\: , \nn \\[1.5mm]
  K_{\,a,2} \Big|_{\,N^{-1}L^2} & \!= \! & \mbox{}
     - 2\,\beta_0^{\,2}\, \cf \:-\: 24\,\beta_0\,\cfs
\:\: , \nn \\[1.5mm]
  K_{\,a,3} \Big|_{\,N^{-1}L^3} & \!= \! & \mbox{}
     - 2\,\beta_0^{\,3}\, \cf \:-\: {88 \over 3}\:\beta_0^{\,2}\,\cfs
\:\: .
\eea

As discussed at the end of Section 3, the vanishing of higher than single-%
enhanced logarithms in $K_{\,a,n}$ leads to relations between coefficient-%
function coefficients at different orders. For the two highest terms at all 
orders one finds 
\bea
\label{tower1}
  d^{\,(n)}_{a,\,2n-1} &\! =\! &
    d^{\,(1)}_{a,\,1} \: \frac{c_{12}^{\,n-1}}{(n-1)!} \;\; \equiv \;\;
    d_{11} \: \frac{c_{12}^{\,n-1}}{(n-1)!} 
    \:\: , 
\\[3mm]
\label{tower2}
  d^{\,(n)}_{a,\,2n-2} &\! =\! &
    d_{11} \, \{ c^{}_{23} - c^{}_{12} c^{}_{11} \} \: 
    \frac{\theta_{n3}\, c_{12}^{\,n-3}}{(n-3)!}
  \:+\: \left\{ d^{\,(2)}_{a,2} - d_{a,0}^{\,(1)} c^{}_{12} \right\} \: 
    \frac{\theta_{n2}\, c_{12}^{\,n-2}}{(n-2)!} 
  \;+\; d_{a,0}^{\,(1)} \: \frac{c_{12}^{\,n-1}}{(n-1)!}
\nn \\[	1mm] &\! =\! &
  d_{11} \left\{ h_{12} \: \frac{\theta_{n3}\, c_{12}^{\,n-3}}{(n-3)!}
  \:+\: h_{21} \: \frac{\theta_{n2}\, c_{12}^{\,n-2}}{(n-2)!} \right\}
  \;+\; d_{a,0}^{\,(1)} \: \frac{c_{12}^{\,n-1}}{(n-1)!} 
  \:\: .
\eea
$d_{a,1}^{\,(1)}$ is independent of $a$, as noted above, hence we denote this
coefficient by $d_{11}$ below. $\,\theta_{kl}$ in (\ref{tower2})
has been defined below Eq.~(\ref{g2n}), and the coefficients $h_{12}$ and 
$h_{21}$ in the last line are given by 
\bea
\label{h12-21} 
 & & 
   h_{12} \:\:= \:\: c^{}_{23} - c^{}_{12} c^{}_{11} \:\:= \:\:
   {1 \over 3}\: \beta_0\: c^{}_{12} 
 \:\: , \nn \\ & & 
   d_{11}\, h_{21} \:\:= \:\: 
   d^{\,(2)}_{a,2} \:-\: d_{a,0}^{\,(1)}\, c^{}_{12} \:\: .
\eea
Here the second identity in the first line arises from the soft-gluon
exponentiation (\ref{CN-res}) together with the LL and NLL expansions 
(\ref{g1n}) and (\ref{g2n}).

A comparison with the tower-expansion \cite{av99} of the soft-gluon resummation
reveals that also Eqs.~(\ref{tower1}) and (\ref{tower2}) correspond to an 
exponential structure which can be written as
\bea
\label{dexp}
 \lefteqn{
  C_a (N) - C_a \Big|_{N^{\,0}\, L^k} \:\: = } 
 \nn \\[0.5mm] & &  {1 \over N} \:
  \Big( \big[ \, d_{11} L + d_{\,a,0}^{\,(1)} \,\big] \,\ar \:+\: 
  \big[ \:\widetilde{\:\!d}_{\,a,1}^{\,(2)}\, L + d_{\,a,0}^{\,(2)} \:\big] 
        \,\ar^{\,2} \:+\; \ldots \Big) 
  \: \exp\, \left\{ L\,h_1(\ar L) + h_2(\ar L) + \ldots^{\,} \right\}
  \:\: , \qquad
\eea
where also the functions $h_k$ are defined in terms of a power expansion, 
\beq
 h_{k}(\ar L) \;\; \equiv \;\; \sum_{k=1} \: h_{kn} \, (\ar L)^n \:\: .
\eeq
Notice that $\widetilde{\:\!d}_{\,a,1}^{\,(2)}$ in Eq.~(\ref{dexp}) is not 
identical to $d_{\,a,1}^{\,(2)}$ in Eq.~(\ref{dcoeffs}) -- the latter quantity
receives a contribution from the expansion of the exponential.
In this notation the third tower of logarithms is given by
\bea
\label{tower3}
  d^{\,(n)}_{a,\,2n-3\,} &\! =\! &
   d_{11} \left\{ 
      \frac{\theta_{n3}\, h_{11}^{\,n-3}}{(n-3)!}\, \Big( h_{22}^{}
      + \frac{1}{2} h_{21}^{\,2} \Big)
    \: + \; \frac{\theta_{n4}\, h_{11}^{\,n-4}}{(n-4)!}\, \Big( h_{13}^{}
      + h_{12}^{} h_{21}^{} \Big) 
    \: + \;
   \frac{\theta_{n5}\, g_{11}^{\,n-5}}{2(n-5)!}\: h_{12}^{\,2} \right\}
 \quad \nn \\[1mm] & & \hspn \mbox{}
    + \; d_{\,a,0}^{\,(1)} 
      \left\{ \frac{\theta_{n2}\, h_{11}^{\,n-2}}{(n-2)!}\, h_{21}^{}
    \: + \; \frac{\theta_{n3}\, h_{11}^{\,n-3}}{(n-3)!}\, h_{12}^{} \right\}
    \: + \;
    \widetilde{\:\!d}_{\,a,1}^{\,(2)} \: 
    \frac{\theta_{n2}\, h_{11}^{\,n-2}}{(n-2)!}
\eea
with $\,h_{11} \,=\, c_{12}^{} \,=\, 2\, C_F$. 

The new coefficient $h_{kn}$ entering Eq.~(\ref{tower3}) (and its lower-%
logarithmic generalizations) can be determined iteratively from fixed-order 
information. The exponentiation (\ref{dexp}) then ensures the vanishing of 
the third-highest (and lower) double-logarithmic contributions to the physical 
kernel at all orders in $\as$. Consequently the conjectured single-logarithmic
large-$x$ enhancement of the physical kernel is equivalent to an exponentiation
in Mellin space beyond the leading $N^{\,0} L^k$ contributions.

All coefficients entering Eqs.~(\ref{tower1}), (\ref{tower2}) and (\ref{tower3})
can be determined from present information. The corresponding coefficients of 
the exponent turn out to be the same for $\Fone$ and $\Ftwo$. They read
\bea
\label{h1kS}
 h_{1k}^{} &\! =\! & g_{1k}^{}  \qquad \mbox{for} \qquad k = 1,\: 2,\: 3
\:\: , \\[1mm]
\label{h21S}
 h_{21}^{} &\! =\! & g_{21}^{} \,+\, {1 \over 2}\, \beta_0 \,+\, 6\,\cf
\:\: , \\[1mm]
\label{h22S}
 h_{22}^{} &\! =\! & g_{22}^{} \,+\, {5 \over 24}\, \beta_0^2 
  \,+\, {17 \over 9}\, \beta_0\,\cf \,-\, 18\,\cfs 
\:\: .
\eea
We conclude that the $1/N$ leading-logarithmic function $h_1(\ar L)$ for DIS is 
identical to its soft-gluon counterpart (\ref{g1n}). 
The function $h_2(\ar L)$, on the other hand, receives additional contributions
which, it appears, prevent direct predictions of $g_{23}^{}$ etc from 
Eqs.~(\ref{h21S}) and (\ref{h22S}). This situation is analogous to that for 
$\FL$ found in Ref.~\cite{MV3}. Hence also here the present predictivity of the
exponentiation is restricted to the three highest logarithms at all higher 
orders in $\as$.
 
The prefactor functions in Eq.~(\ref{dexp}) required to this accuracy are 
given by the coefficient
\beq
\label{d11}
 d_{11} \;\: = \;\: 
 2\,\cf 
%
 \;\; ,
\eeq
and for ${\cal F}_1$ -- and ${\cal F}_{3\,}$, recall the discussion below 
Eq.~(\ref{dcoeffs}) -- by 
\bea
\label{d1-10}
 d_{1,0}^{\,(1)} &\! =\! & 
 {13 \over 2}\:\cf + 2\,\GE\,\cf
%
\\[1mm]
\label{d1-21}
 \widetilde{\:\!d}_{1,1}^{\,(2)} &\! =\! & \mbox{}
  - \cfs \* \left(
    47 + 4\,\z2 - 18\,\GE - 4\,\GEs
    \right)
  + \cf \* \ca \* \Big( 
    {1133 \over 36} - 4\,\z2 + {11 \over 3}\:\GE
    \Big)
\nn \\[0.5mm] & & \mbox{}
  - \cf \* \nf \* \Big( 
    {127 \over 18} + {2 \over 3}\:\GE
    \Big)
%
\:\: .
\eea
The corresponding coefficients for ${\cal F}_2$ read
\bea
\label{d2-10}
 d_{2,0}^{\,(1)} &\! =\! & 
 {21 \over 2}\:\cf + 2\,\GE\,\cf
%
\\[1mm] 
\label{d2-21}
 \widetilde{\:\!d}_{2,1}^{\,(2)} &\! =\! & \mbox{}
  - \cfs \* \left(
    119 - 28\,\z2 - 18\,\GE - 4\,\GEs
    \right)
  + \cf \* \ca \* \Big( 
    {1973 \over 36} - 20\,\z2 + {11 \over 3}\:\GE
    \Big)
\nn \\[0.5mm] & & \mbox{}
  - \cf \* \nf \* \Big( 
    {151 \over 18} + {2 \over 3}\:\GE
    \Big)
%
\:\: .
\eea

Insertion of Eqs.~(\ref{h1kS}) -- (\ref{d2-21}) into Eqs.~(\ref{tower1}), 
(\ref{tower2}) and (\ref{tower3}) provides explicit formulae for the 
coefficients of the three highest $1/N$ logarithms in Eq.~(\ref{dcoeffs}) at
all orders in $\as$. For brevity, we here only present the fourth-order 
results, Mellin-inverted back to $x$-space. For ${\cal F}_1$ one obtains
\bea
\label{c14x}
  c_{1,4}^{}(x) &\!\! =\! & 
  c_{1,4}^{} \Big|_{{\cal D}_k,\delta(1-x)} 
  - {16 \over 3}\: \cff\; L_x^{\,7}
  + \Bigg\{
    {232 \over 3} \:\cff + {28 \over 3} \:\cft\,\beta_0 
  \Bigg\} \: L_x^{\,6}
\nn \\[1mm] & & 
  - \Bigg\{ 
       [ 188 - 128\, \z2 ] \,\cff
    +  [ 12 - 48\,\z2 ] \,\cft \* \ca
    + {1460 \over 9}\: \cft\,\beta_0
    + {16 \over 3}\: \cfs\,\beta_0^{\,2}  
  \Bigg\} \: L_x^{\,5} 
%
\quad \nn \\[1mm] & & 
     \; + \; O(L_x^{\,4}) 
 \:\: ,
\eea
where we have used the abbreviations
$$
  {\cal D}_n \:\; \equiv \;\: [ \x1^{-1}\, \ln^{\,n} \x1 ]_+^{} 
    \quad \mbox{ and } \quad
  L_x \:\: \equiv \:\: \ln \x1
  \:\: .
$$
The coefficients of $\cff\: L_x^{\,7}$, $\cft\:\beta_0\, L_x^{\,6}$ and
$\cfs\,\beta_0^{\,2}\: L_x^{\,5}$ in Eq.~(\ref{c14x}) are the negative of those
of the corresponding +-distributions given (in terms of $C_F$, $C_A$ and 
$\nf$) in Eqs.~(5.4) -- (5.6) of Ref.~\cite{MVV7}. Hence the general pattern 
noted below Eq.~(\ref{pqq0}) is part of the present exponentiation and 
predicted to persist to higher orders.
The corresponding result for ${\cal F}_2$ can be written as
\bea
\label{c24x}
 c_{2,4}^{}(x) &\!\! =\! & 
 c_{1,4}^{}(x)
  + {16 \over 3}\: \cff\: L_x^{\,6}
\nn \\[1mm] & & 
    + \bigg\{
      [ 72 - 64\, \z2 ] \,\cff
    - 32\, [ 1 - \z2 ] \, \cft \* \ca
    - {40 \over 3}\: \cft\, \beta_0
  \bigg\} \: L_x^{\,5} 
%
  \; + \; O(L_x^{\,4}) \:\: . \quad\quad
\eea
This result, obtained from the subleading terms of the physical kernels of 
$\Fone$ and $\Ftwo$, is consistent with Eq.~(16) (which also provides the
coefficient of $L_x^{\,4\,}$) of Ref.~\cite{MV3}, derived from the leading 
large-$x$ physical kernel of the longitudinal structure function $\FL$. This 
agreement provides a rather non-trivial confirmation of our approach.

Although it is not fully known at present, it is instructive to consider
also the fourth tower of logarithms. The corresponding generalization of 
Eq.~(16) of \cite{av99} to the present case (\ref{dcoeffs}) reads
\bea
\label{tower4}
  d^{\,(n)}_{a,\,2n-4\,} &\!\! =\! &
   d_{11} \left\{
      \frac{\theta_{n3}\, h_{11}^{n-3}}{(n-3)!}\: h_{a,3}^{\,(1)} 
    + \frac{\theta_{n4}\, h_{11}^{n-4}}{(n-4)!}\: 
      \Big( h_{23}^{} + h_{22}^{} h_{21}^{} + \frac{1}{6}\, h_{21}^{\,3} \Big)
    + \frac{\theta_{n7}\, h_{11}^{n-7}}{6(n-7)!}\: h_{12}^{\,3} 
 \right. \nn \\ & & \left. \mbox{}
    + \frac{\theta_{n5}\, h_{11}^{n-5}}{(n-5)!}\: 
      \Big( h_{14}^{} + h_{13}^{} h_{21}^{} + h_{12}^{} h_{22}^{} 
      + \frac{1}{2} h_{12}^{} h_{21}^{\,2} \Big)
    + \frac{\theta_{n6}\, h_{11}^{n-6}}{(n-6)!}\: 
      \Big( h_{13}^{} h_{12}^{} + \frac{1}{2}\, h_{12}^{\,2} h_{21}^{} \Big)
 \right\} 
 \nn \\[1mm] & & \hspn \mbox{}
   + \; d_{\,a,0}^{\,(1)} \left\{
      \frac{\theta_{n3}\, h_{11}^{\,n-3}}{(n-3)!}\: \Big( h_{22}^{}
      + \frac{1}{2} h_{21}^{\,2} \Big)
    \: + \; \frac{\theta_{n4}\, h_{11}^{\,n-4}}{(n-4)!}\: \Big( h_{13}^{}
      + h_{12}^{} h_{21}^{} \Big)
    \: + \;
   \frac{\theta_{n5}\, g_{11}^{\,n-5}}{2(n-5)!}\: h_{12}^{\,2} \right\}
 \quad \nn \\[1mm] & & \hspn \mbox{}
    + \; \widetilde{\:\!d}_{\,a,1}^{\,(2)} \:
      \left\{ \frac{\theta_{n3}\, h_{11}^{\,n-3}}{(n-3)!}\: h_{21}^{}
    \: + \; \frac{\theta_{n4}\, h_{11}^{\,n-4}}{(n-4)!}\: h_{12}^{} \right\}
    \;\; + \;\;
    d_{\,a,0}^{\,(2)} \;
    \frac{\theta_{n2}\, h_{11}^{\,n-2}}{(n-2)!}
 \;\; .
\eea
The additional second- and third-order coefficients in Eq.~(\ref{tower4}) are
\bea
\label{d1-20}
 d_{1,0}^{\,(2)} &\! =\! & \mbox{}
  - \cfs \* \left(
    {295 \over 4} + 7\,\z2 - 12\,\z3 - {23 \over 2}\:\GE + 4\,\z2\,\GE
    - 31\:\GEs - 4\:\GE^{\:3} \right)
\nn \\[1mm] & & \mbox{}
  + \cf \* \ca \* \Big( 
    {12419 \over 108} - {35 \over 3}\:\z2 - 20\,\z3 + {781 \over 18}\:\GE
    - 4\,\z2\,\GE + {11 \over 3}\:\GEs \Big)
\nn \\[2mm] & & \mbox{}
  - \cf \* \nf \* \Big( 
    {1243\over 54} - {2 \over 3}\:\z2 + {83 \over 9}\:\GE + {2 \over 3}\:\GEs
    \Big)
%
\:\: , 
\\[3mm]
\label{d2-20}
  d_{2,0}^{\,(2)} &\! =\! & \mbox{}
  - \cfs \* \Big( 
    {431 \over 4} + 47\,\z2 - 60\,\z3 + {49 \over 2}\:\GE - 28\,\z2\,\GE
    - 39\:\GEs - 4\:\GE^{\:3} \Big)
\nn \\[1mm] & & \mbox{}
  + \cf \* \ca \* \Big( 
    {17579 \over 108} + {13 \over 3}\:\z2 - 44\,\z3 + {1333 \over 18}\:\GE
    - 20\,\z2\,\GE + {11 \over 3}\:\GEs \Big)
\nn \\[2mm] & & \mbox{}
  - \cf \* \nf \* \Big( 
    {1699\over 54} - {2 \over 3}\:\z2 + {107 \over 9}\:\GE + {2 \over 3}\:\GEs
    \Big)
%
\eea
and
\bea
\label{h1-31}
  h_{1,3}^{\,(1)} &\! =\! & g_{31}^{} \:+\:
        \cfs \* \Big( 160 - {88 \over 3}\:\z2 - 36\:\GE \Big)
  - \cf \* \beta_0 \* \Big( {116 \over 9} + 2\,\z2 - {34 \over 9}\: \GE \Big)
  + \beta_0^{\,2} \* \Big( {51 \over 16} + {5 \over 12}\:\GE \Big)
\nn \\[2mm] & & \mbox{} + (\ca\!-\!2\,\cf) \* \left\{
        \cf \* \Big( {211 \over 6} - {44 \over 3}\:\z2 \Big)
  + \ca \* \Big( {13 \over 3} - {5 \over 3}\,\z2 \Big)
  - \beta_0 \* \Big( {11 \over 6} + \z2 \Big) \right\}
%
\:\: ,
\\[3mm]
\label{h2-31}
  h_{2,3}^{\,(1)} &\! =\! & h_{1,3}^{\,(1)} \:+\:
   136\,\cfs - {160 \over 9}\: \cf \* \beta_0 + {5 \over 6}\: \beta_0^{\,2}
\nn \\[1mm] & & \mbox{} - (\ca\!-\!2\,\cf) \* \Big\{
      ( 80\,\cf - 8\,\beta_0) \* ( 1 - \z2) 
  + 16\, (\ca\!-\!2\,\cf) \* ( \z3 - \z2 ) \Big\}
%
\eea
with $g_{31}^{}$ given in Eq.~(\ref{g31}). Thus, in contrast to 
Eqs.~(\ref{h21S}) and (\ref{h22S}), the NNLL resummation functions 
$h_{a,3}$ are not the same for $a=1$ and $a=2$, and the deviation of their
leading coefficient from $g_{31}^{}$ involves $\zeta$-functions, including 
$\z3$ in the case of ${\cal F}_2$.  

The only other new coefficient entering Eq.~(\ref{tower4}) at order $\as^{\,4}$
is $h_{23}$. This quantity can be constrained, but not completely fixed, from 
the rather obvious extension of Eqs.~(\ref{KaLLN}) for the 
\mbox{($a$-independent)} leading $1/N$ behaviour of the physical kernel to the 
next order,
\beq
\label{KaLLN2}
  K_{\,a,4} \Big|_{\,N^{-1}L^4} \:\; = \;\: \mbox{}
     - 2\,\beta_0^{\,4}\, \cf \:-\: \xi_{\rm DIS_4}^{}\,\beta_0^{\,3}\,\cfs
  \:\: .
\eeq
The first term on the right-hand-side is fixed by the all-order leading-$\nf$ 
result for $C_a$ \cite{Mankiewicz:1997gz}. Moreover the all-$x$ expressions 
(\ref{Ka3}) strongly suggest that terms with a lower power of $\beta_0$ only
contribute to $K_{\,a,4}(N)$ at higher orders in $1/N$. The consistency of 
Eqs.~(\ref{tower4}) -- (\ref{KaLLN2}) then requires
\beq
\label{h23S}
 h_{23}^{} \:\;=\;\: g_{23}^{} \,+\, {1 \over 8}\, \beta_0^{\,3} 
  \,+\, \left(\, {\xi_{\rm DIS_4}^{} \over 8} - {53 \over 18} \,\right) 
  \beta_0^{\,2} \cf \,-\, {34 \over 3}\, \beta_0\,\cfs \,+\, 72\,\cft
  \:\: .
\eeq
As implied by the notation used above, also this coefficient of the NLL 
resummation function is the same for all structure functions (\ref{Fns-def}). 
The missing information for $\xi_{\rm DIS_4}$ is a next-to-leading 
large-$\nf$ contribution to the fourth-order coefficient function. Since the 
leading large-$\nf$ terms were derived more than ten years ago, and enormous
calculational progress has been made in this time, an extension to the next
order in $\nf$ should be feasible in the near future. We will comment on 
relations between the rational coefficients in Eqs.~(\ref{h21S}), (\ref{h22S}) 
and (\ref{h23S}) below Eq.~(\ref{h23D}). 

The resulting next contribution to Eq.~(\ref{c14x}) reads (recall $L_x \,\equiv
\, \ln \x1$)
\bea
\label{c14L4}
  c_{1,4}^{}\Big|_{\,L_x^4} &\! =\! &
   - \cff \* \Big( {1270 \over 3} + 1424\,\z2 + {400 \over 3}\:\z3 \Big)
   + \cft\,\ca \* \Big( {1576 \over 9} - {1312 \over 3}\:\z2 - 400\,\z3 \Big)
\nn \\[1mm] & & \hspn \mbox{}
   + \cft\,\beta_0 \* \Big( {7583 \over 9} - {520 \over 3}\:\z2 \Big)
   - \cfs\,\cas \* \Big( {46 \over 3} + {20 \over 3}\:\z2 \Big)
   + \cfs\,\ca \* \beta_0 \* \Big( {70 \over 3} - 40\,\z2 \Big)
\quad \nn \\[1mm] & & \hspn \mbox{}
   + \cfs\,\beta_0^{\,2} \* \Big( {\xi_{\rm DIS_4}^{} \over 4} 
   + {277 \over 3} \Big)
   + \cf \* \beta_0^{\,3}
%
\:\: .
\eea
As expected, the coefficient of $\cf \beta_0^{\,3}$ is the negative of the 
corresponding coefficient in Eqs.~(5.7) of Ref.~\cite{MVV7}. The presumed
\mbox{$a$-independent} of $\xi_{\rm DIS_4}$ leads to a definite prediction
for the $\ln^{\,4}\! \x1$ term of the fourth-order longitudinal structure 
function, 
\bea
\label{c24L4}
  c_{2,4}^{}\Big|_{\,L_x^4} &\! =\! & c_{1,4}^{}\Big|_{\,L_x^4}
   + \cff\, ( 32\,\z2 - 160\,\z3 )
   - \cft\,\ca\, ( 8 + 224\,\z2 - 208\,\z3 ) 
   + 12\, \cfs\, \beta_0^{\,2} 
\quad \nn \\[1mm] & & \hspn \mbox{}
   - \cft\,\beta_0 \* \Big( 80 - {352 \over 3}\:\z2 \Big)
   + 64\,\cfs\,\cas \* (\z2 - \z3)
   + {176 \over 3} \* \cfs\,\ca \* \beta_0 \* (1 - \z2) 
%
\:\: .
\eea
This result completes the independent re-derivation of Eq.~(16) in 
Ref.~\cite{MV3}. 

The vanishing of the double-logarithmic $\,N^{-1} \ln^{\,6} N$ contribution 
to $K_{\,a,5}(N)$ fixes the final coefficient in Eq.~(\ref{tower4}),
\beq
\label{h14S}
   h_{14}^{} \:\; = \;\: \frac{1}{5} \: C_F \beta_0^{\,3} 
             \:\; = \;\: g_{14}^{} 
\eeq
where the second equality refers to Eq.~(\ref{g1n}). Thus, up to the presently
unknown number $\xi_{\rm DIS_4}$, the four highest $1/N$ (or $\x1^{0\,}$) 
logarithms for the structure functions (\ref{Fns-def}) are fixed to all 
orders in $\as$. Moreover it appears obvious from Eqs.~(\ref{h1kS}) and 
(\ref{h14S}) that $h_1(\ar L)$ is identical to its soft-gluon counterpart 
$g_1(\ar L)$. 

Instead of working out the corresponding all-order $N$-space formalism at the
next power(s) in $1/N$, we close this section on deep-inelastic scattering by 
presenting the fourth-order extension of Eqs.~(\ref{c11ex}) -- (\ref{c33ex}),
recall the last paragraph of Section~3$\,$:
\bea
\label{c14ex}
 c_{1,4}^{}(x)
  & = & 
        \left( \ln^{\,7} \x1 \; 8/3\: \cff
       - \ln^{\,6} \x1 \; 14/3\: \cft\,\beta_0 
       + \ln^{\,5} \x1 \; 8/3\: \cfs\,\beta_0^{\,2} \right) \* \pqq(x)
  \nn \\[1mm]
  &   & \hspn \mbox{}
      + \ln^{\,6} \x1 \, \left[ \cff \* 
        \left\{ \pqq(x)\, ( - 14 - 68/3\: \H(0) )
                + 4 + 8\, \H(0) - \x1 \* ( 6 + 4\, \H(0) ) \right\} \right] 
  \nn \\[1mm]
  &   & \hspn \mbox{}
      + \ln^{\,5} \x1 \; \left[ \cff \* 
        \left\{ \pqq(x)\, ( - 9 - 8\, \THh(1,0) + 448/3\: \Hh(0,0)
                            + 84\, \H(0) - 64\,\z2 ) + 48\, \THh(1,0)  
  \right. \right.  \nn \\
  &   & \hspp \mbox{} \left. \left.
        - 22 - 96\, \Hh(0,0) - 104\, \H(0) - \x1 \* ( 13 + 24\, 
          \THh(1,0) -48\, \Hh(0,0) - 84\,\H(0) - 16\, \z2) \right\} \right.
  \nn \\[0.5mm]
  &   & \left. \mbox{} + \cft\, \beta_0 \* 
       \left\{ \pqq(x)\, ( 41 + 316/9\, \H(0) ) - 10 - 32/3\:\H(0) 
         + \x1 \* ( 41/3 + 16/3\: \H(0) ) \right\} \right.
  \nn \\[1mm]
  &   & \left. \mbox{} + \cft\, \ca \* 
       \left\{ \pqq(x)\, ( 16 + 8\, \THh(1,0) + 8\, \Hh(0,0) - 24\, \z2 )
       + 4 + \x1 \* ( 28 - 8\, \z2) \right\}
  \right. \nn \\[0.5mm]
  &   & \left. \mbox{} + \cft\, (\ca - 2\,\cf) \:
       \pqq(-x)\, ( 16\, \THh(-1,0) - 8\, \Hh(0,0) ) \right]
%
  \nn \\[1mm]
  &   & \hspn \mbox{}
      + {\cal O} \! \left( \:\!\ln^{\,4} \!\x1 \:\!\right)
\:\: ,
\\[2mm]
\label{c24ex}
 c_{2,4}^{}(x)
  & = &
      c_{1,4}^{}(x) + \ln^{\,6} \x1 \: 16/3\: x\:\cff
  \nn \\[1mm]
  &   & \hspn \mbox{}
      + \ln^{\,5} \x1 \: \left[ \cff \* 
        \left\{ 16 - x\, ( 8 + 48\, \H(0) ) \right\}
      - 40/3\,x\:\cft\, \beta_0
      - \cft\, (\ca\!-\!2\,\cf) \: 32\:\! x \, (1 - \z2) \right]
  \nn \\[1.5mm]
  &   & \hspn \mbox{}
      + \ln^{\,4} \x1 \; \Big[ \cff\,
      \left\{ 8 - 80\,\H(0) - (24 + 48\,\THh(1,0) - 288\,\Hh(0,0)
                       - 48\,\H(0) + 160\,\z2 )\: x \right\}
  \nn \\[1mm]
  &   & \mbox{}
      - \cft\, \beta_0\, \left\{ 112/3 - ( 224/3 + 104\,\H(0) )\: x \right\}
      + \cfs\, \beta_0^{\,2}\: 12\:\!x
  \nn \\[1mm]
  &   & \mbox{} + \cft \* (\ca\!-\!2\,\cf) \* 
      \left\{ - 64/(5\,x^{\,2})\: ( \THh(-1,0) - \z2 /2)
       - 64/(5\,x)\: ( 1 - \H(0) ) - 304/5
  \right. \nn \\ & & \hspp \left. \mbox{}
       - 64\,\THh(-1,0) - 32/5\: \H(0) + 96\,\z2
       + x \, ( 424/5 + 32\, [ 2\,\THhh(-1,-1,0) - \THhh(-1,0,0)
  \right. \nn \\[1mm] & & \hspp \left. \mbox{}
            + \THhh(1,0,0) - \THh(-1,0) + \Hh(0,0) ] 
            + 1168/5\,\H(0) - 80\,\z2 - 192\,\z2\,\H(0) - 48\,\z3)
  \right. \nn \\[1mm] & & \hspp \left. \mbox{}
       - 96/5\: x^{\,2} \* ( 1 + \H(0) )
       + 96/5\: x^{\,3} \* ( \THh(-1,0) - \Hh(0,0) + \z2 /2) \right\} 
  \nn \\[-0.5mm]
  &   & \mbox{}
       + \cfs\, (\ca\!-\!2\,\cf)\, \beta_0 \: 176/3\: x\, ( 1 - \z2) 
       + \cfs\, (\ca\!-\!2\,\cf)^2 \: 64\:\! x\, (\z2 - \z3)  \Big] 
%
  \nn \\[1mm]
  &   & \hspn \mbox{}
      + {\cal O} \! \left( \:\!\ln^{\,3} \!\x1 \:\!\right)
\:\: ,
\\[2mm]
\label{c34ex}
 c_{3,4}^{}(x)
  & = &
      c_{1,4}^{}(x) - \ln^{\,6} \x1 \; 8/3\:\cff \,\x1
  \nn \\[1mm]
  &   & \hspn \mbox{}
      + \ln^{\,5} \x1 \; \left[ \cff \* 
        \left\{ -64\, \H(0) + \x1 \* ( 36 + 56\, \H(0) - 32\,\z2 ) \right\}
  \right.  \nn \\[1.5mm]
  &   & \left. \mbox{}
      + 20/3\:\cft\, \beta_0\, \x1
      + \cft\, \ca \* \left\{ 32\, \H(0) - 16\,\x1 \* ( 1 + \H(0) - \z2)
      \right\}
  \right. \nn \\[0.5mm]
  &   & \left. \mbox{}
      - \cft\, (\ca - 2\,\cf) \,16\,\pqq(-x)\, ( 2\,\THh(-1,0) - \Hh(0,0) )
      \right]
%
 \nn \\
  &   & \hspn \mbox{}
      + {\cal O} \! \left( \:\!\ln^{\,4} \!\x1 \:\!\right)
 \:\: .
\eea
The $\ln^{\,4}\!\x1$ contribution to $c_{1,4}^{}$ involves two unknown 
coefficients of $K_{a,4}(x)$, see Eqs.~(\ref{Ka3}) and (\ref{KaLLN2}). 
The corresponding terms in Eq.~(\ref{c34ex}) can be predicted completely.
However, as $C_3 - C_1\,$ does not correspond to an observable, 
we have refrained from writing them down here.
%
%
\setcounter{equation}{0}
\section{Results for fragmentation and the Drell-Yan process}
\label{sec:largex3}
%
%
As discussed above, the subleading large-$x\,/\,$large-$N$ structure of the 
coefficient functions for the fragmentation functions (\ref{FTns-def}) in 
semi-inclusive $e^+e^-$ annihilation (SIA) is completely analogous to that of 
their DIS counterparts addressed in the previous section. Consequently the 
notation (\ref{dcoeffs}) can be used for the present $1/N$ coefficients as well.
Also these contributions can be resummed in the form (\ref{dexp}), with the 
first four towers of logarithms given by Eqs.~(\ref{tower1}) -- (\ref{tower3}) 
and (\ref{tower4}).

The coefficient functions $C_T$ and $C_A$ are identical up to terms of order
$1/N^2$ or $\x1\,$, cf.~Eqs.\ (\ref{cA1ex}) -- (\ref{cA3ex}) above. The leading
$1/N$ logarithms of the physical kernels (\ref{Fevol}) are the same for all 
three fragmentation functions ${\cal F}_I^{} \,\equiv\, {\cal F}_{T\,}^{} \!+\! 
{\cal F}_{L\,}^{}$, ${\cal F}_T^{}$ and ${\cal F}_{\!A\,}^{}$ 
(recall $L \,\equiv\, \ln N\,$),
\bea
\label{KTaLLN}
  K_{\,a,1} \Big|_{\,N^{-1}L}\; & \!= \! & \mbox{}
     - 2\,\beta_0\, \cf \:+\: 16\,\cfs
\:\: , \nn \\[1mm]
  K_{\,a,2} \Big|_{\,N^{-1}L^2} & \!= \! & \mbox{}
     - 2\,\beta_0^2\, \cf \:+\: 24\,\beta_0\,\cfs
\:\: , \nn \\[1mm]
  K_{\,a,3} \Big|_{\,N^{-1}L^3} & \!= \! & \mbox{}
     - 2\,\beta_0^3\, \cf \:+\: {88 \over 3}\:\beta_0^2\,\cfs
\:\: , \nn \\[1mm]
  K_{\,a,4} \Big|_{\,N^{-1}L^4} & \!= \! & \mbox{}
     - 2\,\beta_0^{\,4}\, \cf \:+\: \xi_{\rm SIA_4}^{}\,\beta_0^{\,3}\,\cfs
\:\: .
\eea
The first three lines derive from Eq.~(\ref{Ka3}). These results are identical
to Eqs.~(\ref{KaLLN}) for the DIS kernels except for the different sign of the
non-leading large-$\nf$ terms. The close relation between the SIA and DIS 
cases suggests $\,\xi_{\rm SIA_4}^{} = \xi_{\rm DIS_4}^{}$ for the fourth-order
generalization in the final line corresponding to Eq.~(\ref{KaLLN2}).

The expansion coefficients of the LL and NLL contributions to the resummation 
exponential (\ref{dexp}), fixed by Eqs.~(\ref{KTaLLN}) and the vanishing of
higher than single-logarithmic contributions, read
\bea
\label{h1kT}
 h_{1k}^{} &\! =\! & g_{1k}^{}  \qquad \mbox{for} \qquad k = 1,\: \ldots, \: 4
\:\: , \\[1mm]
\label{h21T}
 h_{21}^{} &\! =\! & g_{21}^{} \,+\, {1 \over 2}\, \beta_0 \,-\, 6\,\cf
\:\: , \\[1mm]
\label{h22T}
 h_{22}^{} &\! =\! & g_{22}^{} \,+\, {5 \over 24}\, \beta_0^2
  \,-\, {17 \over 9}\, \beta_0\,\cf \,-\, 18\,\cfs
\:\: , \\[1mm]
\label{h23T}
 h_{23}^{} &\! =\! & g_{23}^{} \,+\, {1 \over 8}\, \beta_0^{\,3}
  \,+\, \left( \, {53 \over 18} - {\xi_{\rm SIA_4}^{} \over 8} \,\right)
  \beta_0^{\,2} \cf \,-\, {34 \over 3}\, \beta_0\,\cfs \,-\, 72\,\cft
\:\: .
\eea
The coefficients (\ref{h21T}) -- (\ref{h23T}) differ from their DIS 
counterparts (\ref{h21S}), (\ref{h22S}) and (\ref{h23S}) only by the signs of
every second term in the expansion in powers of $\beta_0$. The first 
coefficients of the NNLL resummation function $h_{a,3}$ (defined as $g_3^{}$ in
Eq.~(\ref{CN-res})), on the other hand, are neither the same for the 
coefficient functions $C_{T,\,A}$ and $C_{I\,}$, the SIA analogue of $C_{2\,}$,
nor do they show a close relation to their DIS counterparts (\ref{h1-31}) and 
(\ref{h2-31}). These coefficients are
\bea
\label{hT-31}
  h_{T,3}^{\,(1)} &\! =\! & g_{31}^{} \:-\:
        \cfs \* \Big( 240 - {88 \over 3}\:\z2 + 36\:\GE \Big)
  - \cf \* \beta_0 \* \Big( {139 \over 9} + 2\,\z2 + {34 \over 9}\: \GE \Big)
  - \beta_0^{\,2} \* \Big( {9 \over 16} - {5 \over 12}\:\GE \Big)
\nn \\[2mm] & & \mbox{} - (\ca\!-\!2\,\cf) \* \left\{
        \cf \* \Big( {1 \over 6} - {44 \over 3}\:\z2 \Big)
  + \ca \* \Big( {34 \over 3} - {5 \over 3}\,\z2 \Big)
  - \beta_0 \* \Big( {49 \over 6} - \z2 \Big) \right\}
%
\:\: ,
\\[3mm]
\label{hI-31}
  h_{I,3}^{\,(1)} &\! =\! & h_{T,3}^{\,(1)} \:+\:
  20\, \cfs + {8 \over 9}\: \cf \* \beta_0 
   + {5 \over 12}\: \beta_0^{\,2}
\nn \\[1mm] & & \mbox{} + (\ca\!-\!2\,\cf) \* \Big\{
      ( 8\,\cf + 4\,\beta_0) \* ( 1 - \z2)
  - 8\, (\ca\!-\!2\,\cf) \* ( \z3 - \z2 ) \Big\}
%
\:\: .
\eea

Finally
the required coefficients of the prefactors of the exponential, again obtained 
by expanding Eq.~(\ref{dexp}) in powers of $\as$ and comparing to the results 
in Section 3, are given by
\beq
\label{d11T}
 d_{11} \;\: = \;\:   
 2\,\cf 
%
\eeq
for both coefficient functions, the same result as in Eq.~(\ref{d11}) for the
DIS case, 
\bea
\label{d1-T0}
 d_{T,0}^{\,(1)} &\! =\! & \mbox{} 
 - {23 \over 2}\,\cf + 2\,\GE\,\cf
%
\\[1mm]
\label{d1-T1}
 \widetilde{\:\!d}_{T,1}^{\,(2)} &\! =\! & \mbox{}
  - \cfs \* \left(
    97 - 20\,\z2 + 6\,\GE - 4\,\GEs
    \right)
  + \cf \* \ca \* \Big( 
    {665 \over 36} - 4\,\z2 + {11 \over 3}\,\GE
    \Big)
 \nn \\[1mm] & & \mbox{}
  - \cf \* \nf \* \Big( 
    {19 \over 18} + {2 \over 3}\,\GE
    \Big)
%
\\[1mm]
\label{d2-T0}
 d_{T,0}^{\,(2)} &\! =\! & \mbox{}
    \cfs \, \Big( 
    {481 \over 4} - 157\,\z2 + 12\,\z3 - {125 \over 2}\:\GE + 20\,\z2\,\GE
    - 29\:\GEs + 4\:\GE^{\:3} \Big)
\nn \\[1mm] & & \mbox{}
  - \cf \* \ca \* \Big( 
    {9325 \over 108} - {37 \over 3}\:\z2 + 20\,\z3 + {47 \over 18}\:\GE
    + 4\,\z2\,\GE - {11 \over 3}\:\GEs \Big)
\nn \\[2mm] & & \mbox{}
  + \cf \* \nf \* \Big( 
    {989\over 54} + {2 \over 3}\:\z2 + {25 \over 9}\:\GE - {2 \over 3}\:\GEs
    \Big)
%
\eea
and
\bea
\label{d1-I0}
 d_{I,0}^{\,(1)} &\! =\! & \mbox{} 
 - {19 \over 2}\,\cf + 2\,\GE\,\cf
%
\\[1mm]
\label{d1-I1}
 \widetilde{\:\!d}_{I,1}^{\,(2)} &\! =\! & \mbox{}
  - \cfs \* \left(
    109 - 36\,\z2 + 6\,\GE - 4\,\GEs
    \right)
  + \cf \* \ca \* \Big( 
    {1085 \over 36} - 12\,\z2 + {11 \over 3}\,\GE
    \Big)
 \nn \\[1mm] & & \mbox{}
  - \cf \* \nf \* \Big( 
    {31 \over 18} + {2 \over 3}\,\GE
    \Big)
%
\\[1mm]
\label{d2-I0}
 d_{I,0}^{\,(2)} &\! =\! & \mbox{}
    \cfs \, \Big( 
    {413 \over 4} - 153\,\z2 + 36\,\z3 - {161 \over 2}\:\GE + 36\,\z2\,\GE
    - 25\:\GEs + 4\:\GE^{\:3} \Big)
\nn \\[1mm] & & \mbox{}
  - \cf \* \ca \* \Big( 
    {6745 \over 108} - {61 \over 3}\:\z2 + 32\,\z3 - {229 \over 18}\:\GE
    + 12\,\z2\,\GE - {11 \over 3}\:\GEs \Big)
\nn \\[2mm] & & \mbox{}
  + \cf \* \nf \* \Big( 
    {761\over 54} + {2 \over 3}\:\z2 + {13 \over 9}\:\GE - {2 \over 3}\:\GEs
    \Big)
%
\:\: .
\eea
Except for the coefficients with $\z3$ (and some obvious terms with $\GE$) 
there is no direct relation either between Eqs.~(\ref{d1-T0}) -- (\ref{d2-I0}) 
and their DIS counterparts (\ref{d11}) -- (\ref{d2-21}), (\ref{d1-20}) and
(\ref{d2-20}).

Inserting Eqs.~(\ref{h1kT}) -- (\ref{d2-I0}) into Eqs.~(\ref{tower1}) --
(\ref{tower3}) and (\ref{tower4}) we arrive at explicit predictions for the
coefficients of the four highest $1/N$ logarithms to all orders in $\as$, with
the fourth logarithm including the unknown coefficient $\,\xi_{\rm SIA_4}$
of Eq.~(\ref{h23T}). After Mellin inversion back to $x$-space the fourth-order
result for ${\cal F}_T$ (and ${\cal F}_{\!A}$, see above) read
\pagebreak
\bea
\label{cT4x}
  c_{T,4}^{}(x) &\!\! =\! &
  c_{T,4}^{} \Big|_{{\cal D}_k,\delta(1-x)}
    - {16 \over 3}\: \cff\; L_x^{\,7}
    + \Bigg\{
    {16 \over 3} \:\cff + {28 \over 3} \:\cft\,\beta_0
  \Bigg\} \: L_x^{\,6}
\nn \\[1mm] & & \hspn \mbox{}
  + \Bigg\{
       [ 104 + 32\, \z2 ] \,\cff
    -  [ 52 - 48\,\z2 ] \,\cft \* \ca
    - {424 \over 9}\: \cft\,\beta_0
    - {16 \over 3}\: \cfs\,\beta_0^{\,2}
  \Bigg\} \: L_x^{\,5}
\quad \nn \\[1mm] & & \hspn \mbox{}
  + \Bigg\{ 
      \cff \* \Big( - {44 \over 3} - 272\,\z2 - {400 \over 3}\:\z3
      \Big)
  + \cft\,\ca \* \Big( 
      {964 \over 9} + {112 \over 3}\:\z2 - 400\,\z3 
      \Big)
\nn \\[1mm] & & 
    - \cft\,\beta_0 \* \Big( 
      {223 \over 9} + {280 \over 3}\:\z2 
      \Big)
 - \cfs\,\cas \* \Big( 
      78 - {20 \over 3}\:\z2 
      \Big)
 + \cfs\,\ca \* \beta_0 \* \Big( 
      {290 \over 3} - 40\,\z2 
      \Big)
\quad \nn \\[1mm] & & 
    + \cfs\,\beta_0^{\,2} \* \Big( 
      {115 \over 3} - {\xi_{\rm SIA_4}^{} \over 4} 
      \Big)
 + \cf \* \beta_0^{\,3}
 \Bigg\} \: L_x^{\,4}
%
     \; + \; O(L_x^{\,3})
 \:\: ,
\eea
where we have again used the abbreviations introduced below Eq.~(\ref{c14x}).
As for the corresponding \mbox{+-distributions}, see Ref.~\cite{MV4}, 
the coefficients of $\cff\: L_x^{\,7}$, $\cft\:\beta_0\, L_x^{\,6}$,
$\cfs\,\beta_0^{\,2}\: L_x^{\,5}$ and $\cf\,\beta_0^{\,3}\: L_x^{\,4}$ in 
Eq.~(\ref{cT4x}) are the same as in Eqs.~(\ref{c14x}) and (\ref{c14L4}) for the
DIS case.

The corresponding predictions for the total fragmentation function 
${\cal F}_I = {\cal F}_T + {\cal F}_L$ lead to the following results for the
longitudinal fragmentation function ${\cal F}_L$:
\bea
\label{cL4x}
 c_{L,4}^{}(x) &\!\! =\! & 
   {8 \over 3}\: \cff\: L_x^{\,6}
   + \bigg\{
      [ 36 - 32\, \z2 ] \,\cff
    - 16\, [ 1 - \z2 ] \, \cft \* \ca
    - {20 \over 3}\: \cft\, \beta_0
  \bigg\} \: L_x^{\,5} 
\nn \\[1mm] & & \hspn \mbox{} + \bigg\{
    \cff\, ( 64\,\z2 - 80\,\z3 )
  - \cft\,\ca\, ( 4 + 112\,\z2 - 104\,\z3 )
  - \cft\,\beta_0 \* \Big( 40 - {176 \over 3}\:\z2 \Big)
\quad \nn \\[1mm] & & \mbox{}
  +  6\, \cfs\, \beta_0^{\,2}
  +  32\,\cfs\,\cas \* (\z2 - \z3)
  +  {88 \over 3} \* \cfs\,\ca \* \beta_0 \* (1 - \z2) \bigg\} \: L_x^{\,4}
%
 \;\: + \;\: O(L_x^{\,3}) \:\: . \quad\quad
\eea
Besides an overall factor of two arising from the different definitions of
${\cal F}_L$ in SIA and DIS, this expression differs from its counterparts 
(\ref{c24x}) and (\ref{c24L4}) for the longitudinal structure functions in 
DIS only in the coefficient of $\z2\, \cff\, \ln^{\,4} \!\x1$.
Eq.~(\ref{cL4x}) can be derived also via the physical evolution kernel for the 
longitudinal fragmentation function, in complete analogy with the DIS case in
Ref.~\cite{MV3}. In fact, Eqs.~(20) -- (22) of that article hold for the 
present case as well, with the above difference arising from the second-order 
prefactor to the resummation exponential. This close relation between the
spacelike and timelike cases does not persist at higher orders in $\x1$, as
can be seen already by comparing Eqs.~(\ref{c21ex}) and (\ref{cL1tex}).


We now turn to the corresponding results for the non-singlet Drell-Yan cross 
section (\ref{DYns-def}). The leading $1/N$ contributions to its physical 
kernel are given by
G%
\bea
\label{KDYLLN}
  K_{\rm DY,1} \Big|_{\,N^{-1}L}\; & \!= \! & \mbox{}
     - 8\,\b0\, \cf \:-\: 32\,\cfs
\:\: , \nn \\[0.8mm]
  K_{\rm DY,2} \Big|_{\,N^{-1}L^2} & \!= \! & \mbox{}
     - 16\,\bb02\, \cf \:-\: 112\,\b0\,\cfs
\:\: , \nn \\[0.8mm]
  K_{\rm DY,3} \Big|_{\,N^{-1}L^3} & \!= \! & \mbox{}
     - 32\,\bb03\, \cf \:+\: \xi_{\rm DY_3}^{}\,\bb02\,\cfs
\:\: , \nn \\[0.8mm]
  K_{\rm DY,4} \Big|_{\,N^{-1}L^4} & \!= \! & \mbox{}
     - 64\,\bb04\, \cf \:+\: \xi_{\rm DY_4}^{}\,\bb03\,\cfs
\:\: .
\eea
Here the first two lines follow from Eqs.~(\ref{KDY2}), while the third and
the fourth are the obvious generalization to order $\as^{\,4}$ and $\as^{\,5}$,
respectively, exploiting the complete analogy to the DIS and SIA cases
discussed above. Also these parts of Eqs.~(\ref{KDYLLN}) are of some interest
despite the unknown subleading large-$\b0$ terms.

This can be seen from the resulting coefficients of the LL and NLL resummation
exponents,  
\bea
\label{h1kD}
 h_{1k}^{} &\! =\! & g_{1k}^{} \qquad \mbox{for} \qquad k = 1,\: \ldots, \: 4
\:\: , \\[2mm]
\label{h21D}
 h_{21}^{} &\! =\! & g_{21}^{} \,+\, \beta_0 \,+\, 7\,\cf
\:\: , \\[1mm]
\label{h22D}
 h_{22}^{} &\! =\! & g_{22}^{} \,+\, {5 \over 6}\, \beta_0^2
  \,-\, \Bigg( 7 - {\xi_{\rm DY_3}^{} \over 24} \Bigg) \beta_0\,\cf
  \,-\, {49 \over 2}\:\cfs
\:\: , \\
\label{h23D}
 h_{23}^{} &\! =\! & g_{23}^{} \,+\, \bb03
  \,-\, \Bigg( \, {7 \over 3} + {\xi_{\rm DY_3}^{} \over 24} 
       - {\xi_{\rm DY_4}^{} \over 32} \Bigg) \bb02 \cf 
  \,-\, \Bigg( 49 - {7\,\xi_{\rm DY_3}^{}  \over 24} \Bigg) \b0\,\cfs 
  \,+\, {343 \over 3}\:\cft
\:\: . \qquad
\eea
We note that, both here and in Eqs.~(\ref{h22S}) and (\ref{h23S}) for the 
structure functions in DIS, the coefficients of $C_F^{\,n}$ in $h_{2n}^{}$ are 
given by $1/n$ times the $n$-th power of the corresponding coefficient 
in~$h_{21}^{}$. Furthermore the coefficients of $\b0 \cfs$ in Eqs.~(\ref{h23S})
and (\ref{h23D}) are the products of the respective $\cf$ and $\b0\cf$ 
coefficients in $h_{21}^{}$ and $h_{22}^{}$. These relations seem to point 
towards a general structure for the functions $h_2(\ar L)$ in Eq.~(\ref{dexp})
which, we hope, can be uncovered in some more deductive approach to the $1/N$ 
contributions to the coefficient functions.

The prefactor coefficients relevant for the highest three logarithms read
\bea
\label{d1-D0}
 d_{\rm DY,1}^{\,(1)} &\! \equiv \! & d_{11} \:\; = \:\; 
8\,\cf
%
 \;\; , \qquad 
 d_{\rm DY,0}^{\,(1)} \:\; = \:\; 
 8\,\GE\,\cf 
%
\;\; ,
\\[1mm]
\label{d1-D1}
 \widetilde{\:\!d}_{\rm DY,1}^{\,(2)} &\! =\! & \mbox{}
  - \cfs \* \left(
    156 - 128\,\z2 - 56\,\GE - 64\,\GEs
    \right)
  + \cf \* \ca \* \Big( 
    {884 \over 9} - 16\,\z2 + {88 \over 3}\,\GE
    \Big)
 \nn \\[0.5mm] & & \mbox{}
  - \cf \* \nf \* \Big( 
    {176 \over 9} + {16 \over 3}\,\GE
    \Big) 
%
\:\: .
\eea
Together with Eqs.~(\ref{h1kD}) -- (\ref{h22D}) these results lead to the
third- and fourth-order predictions
\bea
\label{cD3x}
  c_{\rm DY,3}^{}(x) &\!\! =\! &
  c_{\rm DY,3}^{} \Big|_{{\cal D}_k,\delta(1-x)}
  - 512\, \cft\; L_x^{\,5}
  + \bigg\{
    1728\, \cft + {640 \over 3} \:\cfs\,\beta_0
  \bigg\} \: L_x^{\,4}
\nn \\[1mm] & & \hspn \mbox{}
  + \bigg\{
       [ 2272 + 3072\, \z2 ] \,\cft
    -  \Big[ {544 \over 3} - 512\,\z2 \Big] \,\cfs \* \ca
\nn \\[1mm] & & \mbox{}
    -  \Big[ {2944 \over 3} + {\xi_{\rm DY_3}^{} \over 3} \Big]\: 
       \cfs\,\beta_0
    - {64 \over 3}\: \cf \* \beta_0^{\,2} \bigg\} \: L_x^{\,3}
%
  \:\: + \:\: O(L_x^{\,2})
\eea
and
\bea
\label{cD4x}
  c_{\rm DY,4}^{}(x) &\!\! =\! &
  c_{\rm DY,4}^{} \Big|_{{\cal D}_k,\delta(1-x)}
  -  {4096 \over 3}\: \cff\; L_x^{\,7}
  +  \bigg\{
    {19712 \over 3}\: \cff + {3584 \over 3} \:\cft\,\beta_0
  \bigg\} \: L_x^{\,6}
\nn \\[1mm] & & \hspn \mbox{}
  + \bigg\{
       [ 9088 + 20480\, \z2 ] \,\cff
    -  [ 1408 - 3072\,\z2 ] \,\cft \* \ca
\nn \\[1mm] & & \mbox{}
    -  \Big[ {20864 \over 3} + {8\,\xi_{\rm DY_3}^{} \over 3} \Big]\:
       \cft\,\beta_0
    - {1024 \over 3}\: \cfs\, \beta_0^{\,2} \bigg\} \: L_x^{\,5}
%
  \:\: + \:\: O(L_x^{\,4}) \:\: ,
\eea
where the respective third logarithms depend on the presently unknown 
quantity $\xi_{\rm DY_3}$. Also in Eqs.~(\ref{cD3x}) and (\ref{cD4x}) the 
coefficients of the highest +-distributions and powers of $L_x \equiv \ln \x1$ 
for each colour factor are equal in magnitude but opposite in sign.

Finally we provide the generalizations of Eqs.~(\ref{cT1ex}) -- (\ref{cD2ex}) 
to the next order in $\as$. For the fragmentation functions (\ref{FTns-def}) 
these are given by
\bea
\label{cT4ex}
 c_{T,4}^{}(x)
  & = &
      c_{1,4}^{}(x) + \ln^{\,6} \!\x1 \:
      \cff\, \left\{ 32\,\pqq(x)\,\H(0) - 8 + 4\,\x1 \right\}
  \nn \\[0.5mm]
  &   & \hspn \mbox{}
      + \ln^{\,5} \x1 \; \left[ \cff \* 
        \big\{ \pqq(x)\, ( 16\,\THh(1,0) - 152\,\Hh(0,0) - 108\,\H(0)
                + 48\,\z2 ) \right. 
  \nn \\[1mm]
  &   & \left. \hspp \mbox{}
        + 44 + 176\,\Hh(0,0) + 24\,\H(0)
        - \x1 \* ( 22 + 88\,\Hh(0,0) + 68\,\H(0)  )
      \big\}
  \right.  \nn \\[2mm]
  &   & \left. + \cft\, \beta_0 \* 
        \left\{ -428/9\:\pqq(x)\,\H(0) + 20 - 10\,\x1 \right\}
                \right.
  \nn \\[1mm]
  &   & \left. + \cft\, \ca \* 
        \left\{ \pqq(x)\, ( - 16\,\THh(1,0) - 8\,\Hh(0,0) ) - 8 + 4\,\x1
      \right\} \right]
%
  \nn \\[1mm]
  &   & \hspn \mbox{}
      + {\cal O} \! \left( \:\!\ln^{\,4} \!\x1 \:\!\right)
  \:\: .
\\[2mm]
\label{cL4tex}
 c_{L,4}^{}(x)
  & = &
      \ln^{\,6} \x1 \: 8/3\:\cff
  \nn \\[1mm]
  &   & \hspn \mbox{}
      + \ln^{\,5} \x1 \; \left[ \cff \* 
        \left\{ - 4 + 8\,\H(0) + 8\:\! x \right\}
      - 20/3\:\cft\, \beta_0
      - 16\,\cft \* (\ca\!-\!2\,\cf) \* ( 1 - \z2 ) \right]
  \nn \\[1.5mm]
  &   & \hspn \mbox{}
      + \ln^{\,4} \x1 \; \Big[ \cff\,
      \left\{ 4 - 8\,\THh(1,0) - 8\,\Hh(0,0) - 4\, \H(0) - 32\,\z2 
              - x\, ( 12 - 16\,\H(0) ) \right\}
  \nn \\[1mm]
  &   & \mbox{}
      + \cft\, \beta_0\, \left\{ 112/3 - 56/3\:\H(0) - 56/3\: x \right\}
  \nn \\[1mm]
  &   & \mbox{} + \cft \* (\ca\!-\!2\,\cf) \* 
      \left\{ - 48/(5\,x^{\,2})\: ( \THh(-1,0) - \z2 /2)
       - 48/(5\,x)\: ( 1 - \H(0) ) + 212/5
  \right. \nn \\ & & \hspp \left. \mbox{}
       - 16\, [ 2\,\THhh(-1,-1,0) - \THhh(-1,0,0) - 2\, \THhh(0,-1,0)
       - \THhh(1,0,0) - \THh(-1,0) ] - 184/5\: \H(0) - 40\,\z2
  \right. \nn \\[1mm] & & \hspp \left. \mbox{}
       - 24\,\z3 + 16\,\z2\, \H(0) + x \, ( -152/5 + 32\, \THh(-1,0) 
       - 32\,\Hh(0,0) + 16/5\,\H(0) + 48\,\z2 ) 
  \right. \nn \\[1mm] & & \hspp \left. \mbox{}
       - 32/5\: x^{\,2} \* ( 1 + \H(0) )
       + 32/5\: x^{\,3} \* ( \THh(-1,0) - \Hh(0,0) + \z2 /2) \right\}
  \nn \\[-0.5mm]
  &   & \mbox{}
       + 6\, \cfs\, \beta_0^{\,2}
    + \cfs \* (\ca\!-\!2\,\cf)\, \beta_0 \: 88/3\: ( 1 - \z2)
    + \cfs \* (\ca\!-\!2\,\cf)^2 \: 32\, (\z2 - \z3) \Big] \quad
%
  \nn \\[1mm]
  &   & \hspn \mbox{}
      + {\cal O} \! \left( \:\!\ln^{\,3} \!\x1 \:\!\right)
  \:\: .
\\[2mm]
\label{cA4ex}
 c_{A,4}^{}(x)
  & = &
      c_{T,4}^{}(x) - \ln^{\,6} \x1 \: 8/3\: \cff \,\x1
  \nn \\[1mm]
  &   & \hspn \mbox{}
      + \ln^{\,5} \x1 \; \left[ \cff \* 
        \left\{ -64\, \H(0) + \x1 \* ( 36 + 24\, \H(0) - 32\,\z2 ) \right\}
  \right.  \nn \\[1.5mm]
  &   & \left. \mbox{}
      + 20/3\: \cft\, \beta_0\, \x1
      + \cft \* \ca \* \left\{ 32\, \H(0) - 16\,\x1 \* ( 1 + \H(0) - \z2)
      \right\}
  \right. \nn \\[0.5mm]
  &   & \left. \mbox{}
      - \cft\, (\ca - 2\,\cf) \,16\,\pqq(-x)\, (2\,\THh(-1,0) - \Hh(0,0) )
      \right]
%
  \nn \\
  &   & \hspn \mbox{}
      + {\cal O} \! \left( \:\!\ln^{\,4} \!\x1 \:\!\right)
  \:\: .
\eea
The corresponding result for the third-order Drell-Yan coefficient function 
reads
\bea
\label{cD3ex}
 c_{\rm DY,3}^{}(x)
  & = &
        \left( \ln^{\,5} \x1 \; 192\, \cft
      - 80\, \ln^{\,4} \x1 \, \cfs\, \beta_0 \right) \, \pqq(x)
  \nn \\[0.5mm]
  &   & \hspn \mbox{}
      + \ln^{\,4} \x1 \, \left[ \cft \,
        \left\{ - 648\,\pqq(x)\, \H(0)
                + 384\, \H(0) - 192\,\x1 \* ( 2 + \H(0) ) \right\} \right]
%
 \nn \\[1.5mm]
  &   & \hspn \mbox{}
      + {\cal O} \! \left( \:\!\ln^{\,3} \!\x1 \:\!\right)
  \:\: .
\eea
Unlike Eq.~(\ref{KDYLLN}), the fourth-order generalization of Eqs.~(\ref{KDY2})
involves more than one unknown coefficient, hence we have not included the 
incomplete $\ln^{\,3} \!\x1$ contribution in Eq.~(\ref{cD3ex}).
%
%
\setcounter{equation}{0}
\section{Numerical illustrations}
\label{sec:numerics}
\begin{figure}[p]
\centerline{\epsfig{file=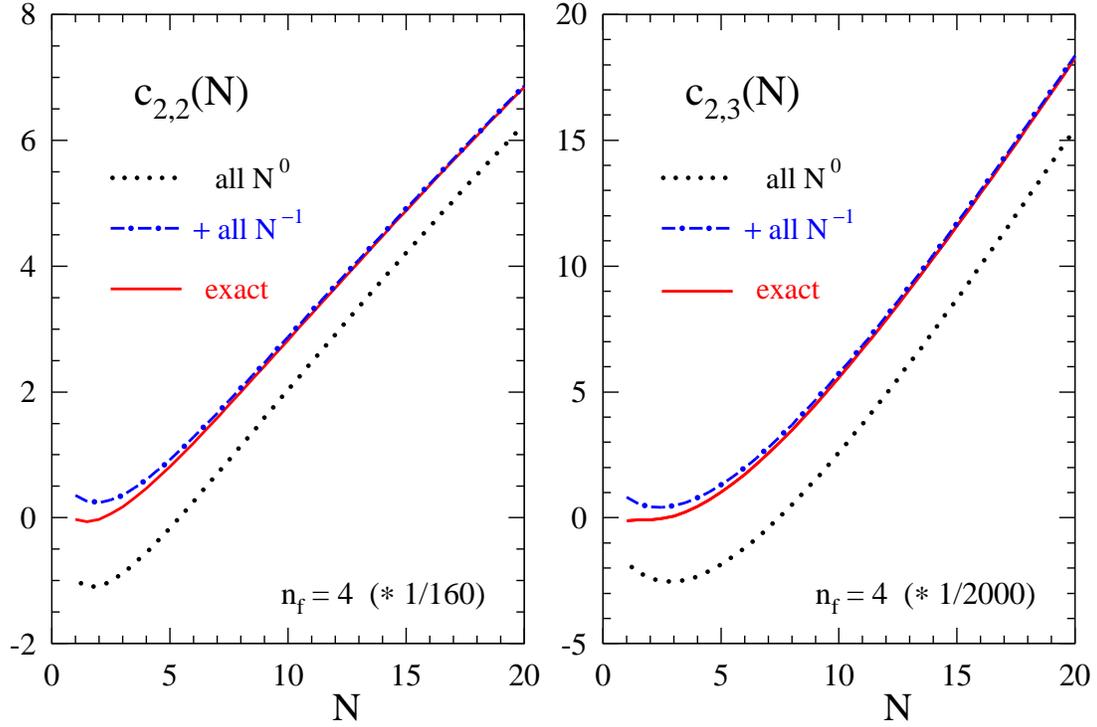,width=15.0cm,angle=0}}
\vspace{-4mm}
\caption{\label{pic:fig1}
 The second- and third-order non-singlet coefficient functions for the structure
 function $\Ftwo$ as defined in Eq.~(\ref{Fns-cq}) in Mellin-$N$ space. 
 The leading and subleading large-$N$ contributions (\ref{c22num}) and
 (\ref{c23num}) are compared to the exact functions for $\nf=4$ light flavours. 
 The results are multiplied by suitable factors compensating our small choice 
 $\ar = \as/(4\pi)$ of the expansion parameter.
 }
\vspace{-1mm}
\end{figure}

%
%
We close by briefly illustrating the numerical size of the known and new 
subleading large-$N$ contributions to the coefficient functions. For $\nf = 4$
the corresponding expansions of the two- and three-loop coefficient function 
for $\Ftwo\,$, the practically most important structure function, are given~by
\bea
\label{c22num}
  c_{2,2}^{}(N) & \!=\! &
       3.556\,L^4 + 26.28\,L^3 + 40.76\,L^2 - 67.13\,L - 157.3
  \nn \\[0.5mm]
  &   & \mbox{}
       + N^{\,-1} ( \,
       7.111\,L^3 + 92.76\,L^2 + 239.5\,L + 214.0 \, )
       \; + \; {\cal O} \left( N^{\,-2\,} \right)
  \:\: ,
\\[2mm]
\label{c23num}
  c_{2,3}^{}(N) & \!=\! & 
       3.160\,L^6 + 44.92\,L^5 + 238.9\,L^4
       + 470.8\,L^3 - 620.2\,L^2 - 1639\,L - 3586
  \nn \\[0.5mm]
  &   & \mbox{} 
       + N^{\,-1} ( \, 
       9.481\,L^5 + 211.9\,L^4 + 1393\,L^3 + 4157\,L^2 + 5200\,L + 5230 \, ) 
  \nn \\[0.5mm]
  &   & \mbox{} 
       + {\cal O} \left( N^{\,-2\,} \right) 
  \:\: . 
\eea
In Fig.~\ref{pic:fig1} these approximations, with and without the $1/N$ terms, 
are compared to the exact results of Refs.~\cite{ZvNcq2,Moch:1999eb} and 
\cite{MVV6}. At both orders the latter contributions are relevant over the full
range of $N$ shown in the figure, while terms of order $1/N^2$ are sizeable
only at $N < 5$. Note that the classification as $N^{\,0}$ and $N^{\,-1}$ terms
does not reflect the numerical behaviour for the $N$-values of the figure. 
E.g., the third-order increase due to the $\ln^{\,k} N$ contributions in the 
first line of Eq.~(\ref{c22num}) strongly resembles a linear rise, and the sum 
of the $N^{\,-1} \ln^{\,k}\! N$ terms in the second line almost looks like a 
constant. In fact, the decrease of this contribution towards large $N$ is very 
slow: only at $N = 1.5\cdot 10^{\,2}$ has it fallen to half of the value at 
its maximum at $N = 6.6$. 
The situation for the corresponding third-order coefficient functions for 
$\Fone$ and $\F3$ \cite{MVV10}, not shown here for brevity, is similar except 
at small $N$ where in both cases the sum of the $N^{\,0}$ and $N^{\,-1}$ terms 
is close to the exact result even down to $N=1$, the lowest value of $N$ used
in the figures.

\begin{figure}[p]
\centerline{\epsfig{file=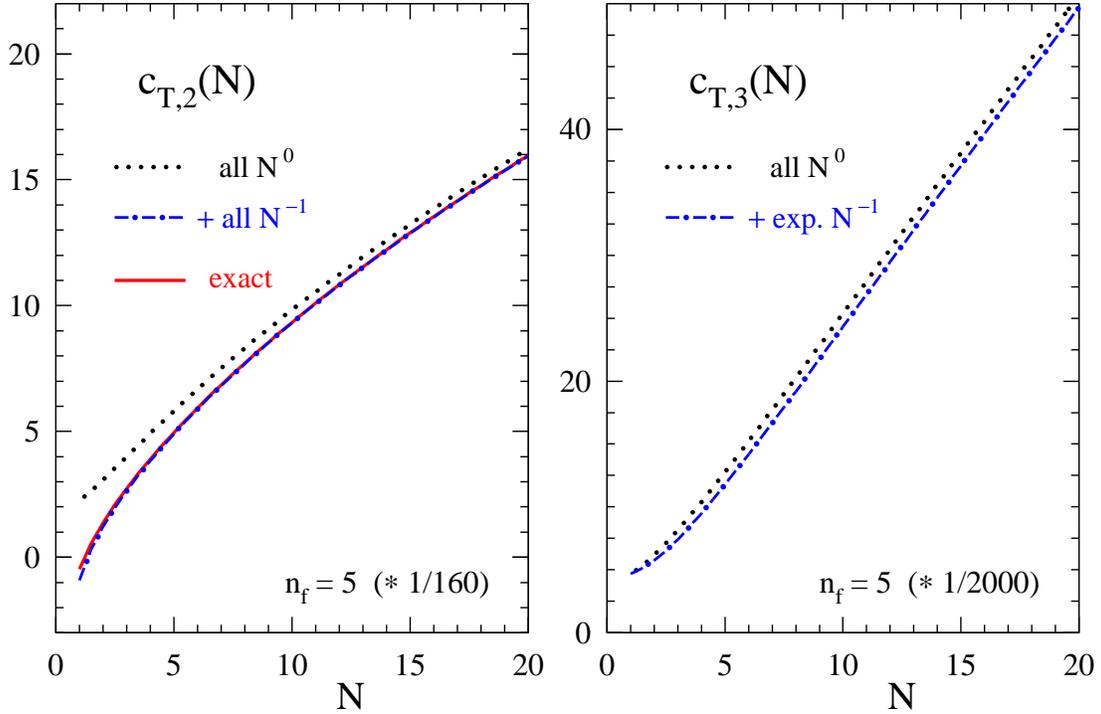,width=15.0cm,angle=0}}
\vspace{-4mm}
\caption{\label{pic:fig2}
 As Fig.~\ref{pic:fig1}, but for the fragmentation function $F_{\:\!T}$ at 
 $\nf = 5$. Neither the exact three-loop result nor the corresponding 
 coefficients of the $N^{\,-1} \ln\, N$ and $N^{\,-1}$ terms are known at 
 present.
 }
\vspace{-1mm}
\end{figure}

The pattern of the coefficients is rather different for both the $N^{\,0}$ and 
$N^{\,-1}$ contributions to the corresponding coefficient functions for the 
transverse fragmentation function $F_{\:\! T}$,
\bea
\label{cT2num}
  c_{T,2}^{}(N) & \!=\! &
       3.556\,L^4 + 25.69\,L^3 + 105.6\,L^2 + 104.3\,L 
  \nn \\[0.5mm]
  &   & \mbox{}
       + N^{\,-1} ( \,
       7.111\,L^4 - 29.02\,L^4 - 111.4\,L - 504.0\, )
       + {\cal O} \left( N^{\,-2\,} \right)
 \:\: ,
\\[2mm]
\label{cT3num}
  c_{T,3}^{}(N) & \!=\! &
       3.160\,L^6 + 43.34\,L^5 + 309.3\,L^4
       + 1017\,L^3 + 2306\,L^2 + 2090\,L + 9332
  \nn \\[0.5mm]
  &   & \mbox{}
       + N^{\,-1} ( \,
       9.481\,L^5 - 10.17\,L^4 - 362.7\,L^3 - 3247\,L^2 \, )
       + {\cal O} \left( N^{\,-1}\, L \right)
 \:\: .
\eea
These expansions are shown in Fig.~\ref{pic:fig2} together with the exact
second-order result of Refs.~\cite{RvN96,Mitov:2006wy}.
As adequate for an observable measured in particular at scales not too far from 
the $Z$-mass, the results refer to $\nf = 5$ effectively light flavours. 
All $N^{\,0}$ contributions are positive in Eqs.~(\ref{cT2num}) and 
(\ref{cT3num}), yielding a larger soft-gluon enhancement than in the DIS case 
especially due to the lower powers of $\ln N$ as already discussed in 
Ref.~\cite{MV4}. 
On the other hand, the $N^{\,-1} \ln^{\,k}\! N$ coefficients change sign here,
again in contrast to Eqs.~(\ref{c22num}) and (\ref{c23num}).
This leads to smaller and negative $1/N$ corrections which do not exceed 10\% 
except for $N < 7$ in the two-loop case in the left part of the figure, where 
their inclusion results in a good approximation down to $N \simeq 2$. 
At the third order the $N^{\,-1} \ln N$ and $N^{\,-1}$ contributions are not 
yet known. One may expect similarly relevant \mbox{small-$N$} corrections from 
these terms to the corresponding curve shown in the right part of the figure.
Similar results are found for the integrated and asymmetric fragmentation 
functions. 


We now turn to the four-loop predictions derived in the previous two sections, 
again focusing on the same two DIS and SIA observables. The known and new 
contributions at this order read
\bea
\label{c24num}
  c_{2,4}^{}(N) & \!\!=\!\! &
       2.107\,L^8 + 48,71\,L^7 + 477.9\,L^6
       + 2429\,L^5 + 5240\,L^4 - 1824\,L^3- 30308\,L^2
  \nn \\[0.5mm]
  &   & \mbox{}
       + {\cal O} \left( L \right)
       \; + \; N^{\,-1} ( \,
       8.428\,L^7 + 284.3\,L^6 + 3324\,L^5 + 
       [ 18884 + 30.86\, \xi_{\rm DIS_4}^{} ] \,L^4 )
  \nn \\[1.5mm]
  &   & \mbox{}
       + {\cal O} \left( N^{\,-1} L^3 \,\right)
\eea
and
\bea
\label{cT4num}
  c_{T,4}^{}(N) & \!\!=\!\! &
       2.107\,L^8 + 46.60\,L^7 + 514.1\,L^6
       + 3126\,L^5 + 11774\,L^4 + 23741\,L^3 + 46637\,L^2
  \nn \\[0.5mm]
  &   & \mbox{}
       + {\cal O} \left( L \right)
       \; + \; N^{\,-1} ( \,
       8.428\,L^7 + 32.47\,L^6 - 448.1\,L^5 -
       [ 7315 + 26.12 \, \xi_{\rm SIA_4}^{} ] \,L^4 )
  \nn \\[1.5mm]
  &   & \mbox{}
       + {\cal O} \left( N^{\,-1} L^3 \,\right) 
 \:\: .
\eea
As above, the results for $\Ftwo$ in Eq.~(\ref{c24num}) are given for $\nf= 4$, 
and those for $F_{\:\!T}$ in Eq.~(\ref{cT4num}) for $\nf = 5$. The $N^{\,0}$ 
coefficients have been presented already in Tables 1 of Refs.~\cite{MVV7,MV4}. 
The $\ln^{\,2}N$ term in both equations includes a small contribution $A_4/2$ 
from the fourth-order cusp anomalous dimension for which we have used the 
respective Pad\'e estimates of 4310 for $\nf = 4$ and 1550 for $\nf = 5$ 
\cite{MVV7}.
The fourth $N^{\,-1}$ logarithms receive small contributions from the presently
unknown (and most likely identical) fourth-order coefficients $\xi_{\rm DIS_4}$
and $\xi_{\rm SIA_4}$ of Eqs.~(\ref{KaLLN2}) and (\ref{KTaLLN}). Values expected
from the latter equations contribute less than 2\% to the coefficients of 
$N^{\,-1} \ln^{\,4} N$.

\begin{figure}[p]
\centerline{\epsfig{file=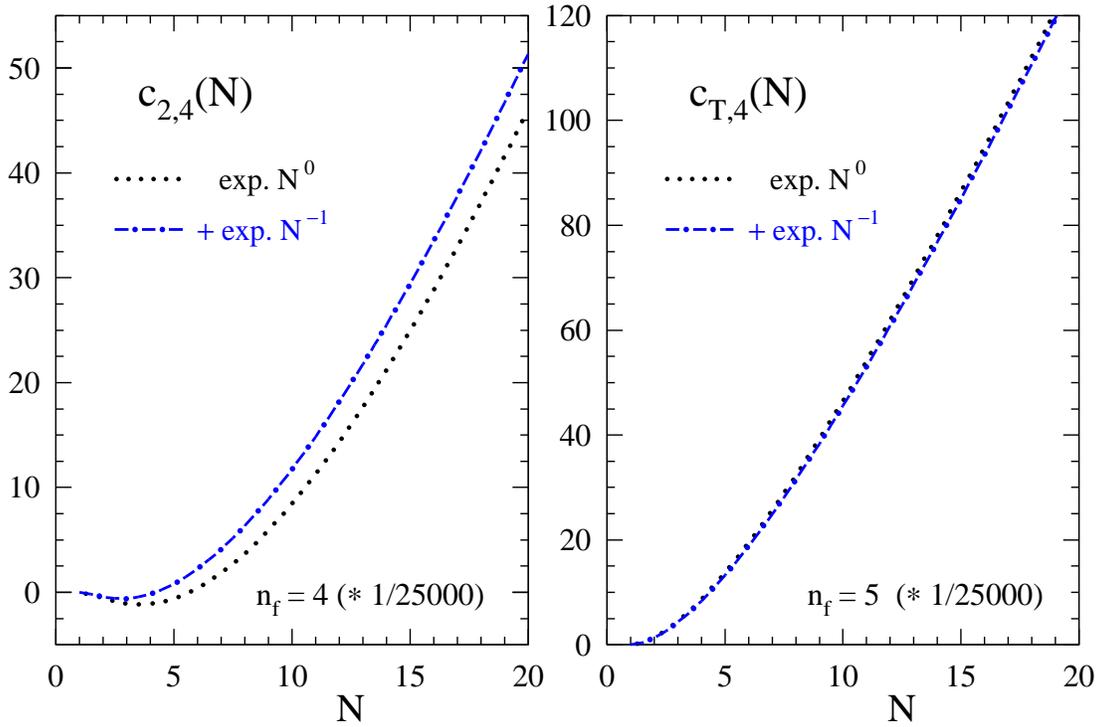,width=15.0cm,angle=0}}
\vspace{-4mm}
\caption{\label{pic:fig3}
 Large-$N$ contributions to the fourth-order non-singlet coefficient functions
 for $\Ftwo$ in DIS (left) and $F_{\:\!T}$ in SIA (right). Shown are the known
 $N^{\,0}$ and $N^{\,-1}$ contributions as given in Eqs.~(\ref{c24num}) and
 Eqs.~(\ref{cT4num}). The results have been multiplied by 25000 $\,\simeq
 (4\pi)^4$ for display purposes.
 }
\vspace{-1mm}
\end{figure}

The presently unknown lower-$k$ $N^{\,-1} \ln^{\,k} N$ terms can be expected to 
enhance the $1/N$ effects shown in Fig.~\ref{pic:fig3}. Yet already now 
one can conclude that the pattern of the previous two orders appears to persist
to order $\as^{\,4}$, e.g., that the $N^{\,-1}$ contributions are small for
$F_{\:\!T}$ at least at $N \gsim 10$.
We stress that this figure does not intend to present the best approximation to
dominant $N^{\,0}$ contributions, but simply illustrates the effect of the
known terms as given in Eqs.~(\ref{c24num}) and (\ref{cT4num}). Rough estimates 
of the missing coefficient of $\ln N$ can be obtained by expanding the 
soft-gluon exponential (\ref{CN-res}) or (also for the non-logarithmic 
$N^{\,0}$ terms) via the Mellin transform of the known seven +-distributions 
given in Eqs.~(5.4) -- (5.10) of Ref.~\cite{MVV7} -- note that there are some
typos in the first archive and journal versions of this article -- 
and in Eq.~(32) of Ref.~\cite{MV4}. The latter article includes also the 
$c_{2,4}^{} - c_{T,4}^{}$ difference of the $\ln N$ and $N^{\,0}$ coefficients
in Eqs.~(\ref{c24num}) and (\ref{cT4num}).

\begin{figure}[p]
\centerline{\epsfig{file=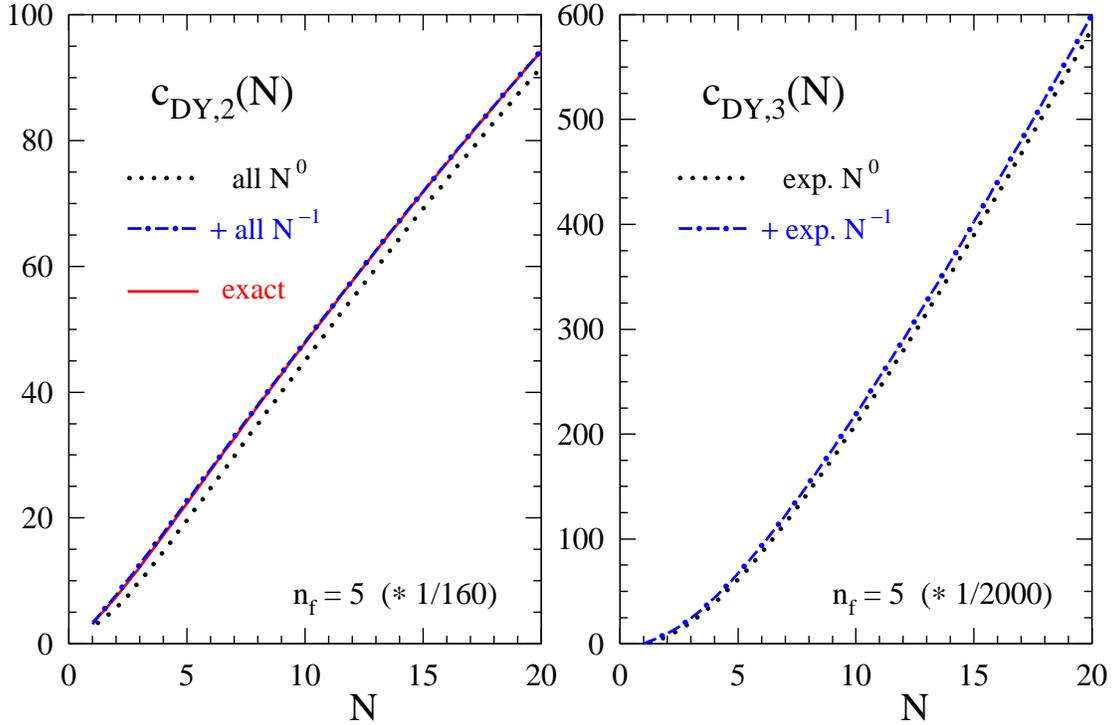,width=15.0cm,angle=0}}
\vspace{-4mm}
\caption{\label{pic:fig4}
 As Fig.~\ref{pic:fig2}, but for Drell-Yan cross section (\ref{DYns-def}),
 using the expansions (\ref{cDY2num}) and (\ref{cDY3num}) with $\xi_{\rm DY_3} 
 = -\,400$.  Besides the coefficients mentioned in the caption of that figure, 
 also the third-order constant-$N$ and $N^{\,-1}\ln^{\,2} N$ coefficient are 
 unknown in this case.
 }
\vspace{-1mm}
\end{figure}

Finally the large-$N$ expansion of the second- and third-order coefficient
functions for the non-singlet (quark-antiquark annihilation) Drell-Yan cross 
section (\ref{DYns-def}) are given by
\bea
\label{cDY2num}
  c_{2,2}^{}(N) & \!=\! &
       56.89\,L^4 + 185.9\,L^3 + 428.6\,L^2 + 267.6\,L + 442.8
  \nn \\[0.5mm]
  &   & \mbox{}
       + N^{\,-1} ( \,
       113.8\,L^3 + 378.4\,L^2 + 577.3\,L + 53.43 \, )
       + {\cal O} \left( N^{\,-2\,} \right)
 \:\: ,
\\[2mm]
\label{cDY3num}
  c_{2,3}^{}(N) & \!=\! &
       202.3\,L^6 + 1282\,L^5 + 4676\,L^4
       + 8172\,L^3 + 11404\,L^2 + 6395\,L + {\cal O} \left( 1 \right)
  \nn \\[0.5mm]
  &   & \mbox{}
       + N^{\,-1} ( \,
       606.8\,L^5 + 4267\,L^4 + [ 12164 - 4.543\,\xi_{\rm DY_3}^{} \, ]\,L^3 )
       + {\cal O} \left( N^{\,-1} L^2 \,\right) 
\eea
for $\nf = 5$. These expansions are shown in Fig.~\ref{pic:fig4} together with
the exact two-loop results of Refs.~\cite{Hamberg:1991np,Harlander:2002wh}.
The higher-order corrections are much larger in this case than in DIS and SIA.
Also here the $1/N$ contributions appear to be numerically rather unimportant, 
a feature that appears to persists to even lower values of $N$ than for the 
fragmentation functions.

%
%
\setcounter{equation}{0}
\section{Summary and outlook}
\label{summary}
%
%
We have analysed the $\ln^{\,k}\!\x1$ contributions to the physical evolution
kernels for -- including the results already presented in Ref.~\cite{MV3} -- 
nine flavour non-singlet observables in inclusive DIS, semi-inclusive $e^+e^-$
annihilation (SIA) and Drell-Yan lepton-pair production. It turns out that
all these kernels include only single-logarithmic higher-order corrections, up
to $\as^{\,n} \x1^k \ln^{\,n-1}\!\x1$, at all powers $k$ of $\x1$. 
On the other hand, the coefficient functions from which these kernels are 
constructed received double-logarithmic contributions up to $\as^{\,n} \x1^k 
\ln^{\,2n-1}\!\x1$ at all orders. This difference implies that the terms 
$\as^{\,n} \x1^k \ln^{\,l}\!\x1$ with $ n \leq l < 2n$ are functions of 
lower-order terms, i.e., a general resummation of the double-logarithmic terms 
at all powers of $\x1$.

The above pattern is established to all order in $\as$ by the soft-gluon
exponentiation of the $\x1^{-1} \ln^{\,k}\!\x1$ contributions to the 
coefficient functions \cite{SoftGlue,MVV7,MV1,Laenen:2005uz,Idilbi:2005ni,%
Blumlein:2006pj,MV4}. All-order results underpinning it at all powers in 
$\x1$ are presently known only for the leading large-$\nf$ contributions to
DIS structure functions \cite{Gracey:1995aj,Mankiewicz:1997gz}. 
However, all available fixed-order results on higher-order coefficient 
functions \cite{ZvNcq2,Moch:1999eb,MVV2,MVV5,MVV6,MVV10,RvN96,Mitov:2006wy,%
Hamberg:1991np,Harlander:2002wh} are consistent with the behaviour described in
the previous paragraph. 

In our view it is most unlikely that this consistency is accidental, given the 
large number of observables and the depth of the perturbative expansion reached
especially in DIS and SIA -- for the latter this article includes some new 
third-order results. 
Moreover it should be noted that the resummation of $\FL$ in both DIS and SIA 
can be consistently constructed each via two different physical kernels: that 
for these quantities themselves (starting with $\x1^{-1\,}$) and via the 
difference (starting with $\x1^{0\,}$) of the respective kernels, $K_2 - K_1$
and $K_I - K_T$, for the structure functions $F_{1,2}$ and fragmentation 
functions $F_{\:\!T,I}$ where $F_{\:\!I}$ is our notation for the total 
(angle-integrated) fragmentation function. 
We thus definitely expect that we are observing a genuine feature of the 
coefficient functions and expect that a more deductive approach, such as that 
pursued in Ref.~\cite{Laenen:2008gt}, can provide a formal proof in the near 
future at least for the next power in $\x1$.

We have employed the conjectured single-logarithmic enhancement of the 
physical kernels to derive the explicit $x$-dependence of the coefficients of 
the three highest powers of the fourth-order DIS and SIA coefficient functions,
while in the Drell-Yan case we are restricted to two logarithms at order 
$\as^{\,3}$. For this purpose we have employed a modified basis (required far 
beyond the weight-3 functions shown in the article) for the harmonic 
polylogarithms. An extension of these results to higher orders in $\as$ is 
possible but not necessary at present in view of the discussion given below. 

For the subdominant (except for $\FL$) logarithms with prefactor $\x1^0$ we
have cast our results in the form of an exponentiation, akin to that of the
$\x1^{-1}$ soft-gluon effects, in Mellin-$N$ space where these terms behave as
$N^{-1} \ln^{\,k} N$. One more logarithm can be effectively predicted in this
case, as the one unknown parameter turns out to be numerically suppressed.
Our resummation of the $1/N$ terms is, nevertheless, far less predictive than 
the soft-gluon exponentiation (which predicts seven of eight fourth-order 
logarithms in DIS and SIA) for two main reasons: Firstly the prefactor of the
exponential is of first instead of zeroth order in $\as$, thus one more order
needs to be calculated in order to fix the same number of coefficients.
Secondly, while the leading-logarithmic function, usually denoted by 
$g_1^{}(\as \ln N)$, in the exponent is the same as in the $N^{\,0}$ soft-gluon 
case, this does not hold for the higher-logarithmic functions which have an
(at least presently) not fully predictable power expansion (from $g_{2\,}^{}$)
and do not show any universality (from $g_{3\,}^{}$).

Finally we have illustrated the numerical size of the $1/N$ contributions.
It turns out that, in the restricted $N$-region of practical interest, the
logarithms at the third and higher orders essentially compensate one power of
$N$, i.e., the $N^{\,0}$ terms together resemble a linear increase with $N$, 
and $1/N$ corrections almost look like a constant. The sum of all $N^{\,0}$ and 
$N^{-1}$ contributions is found to provide an excellent approximation of the
exact results, except at small $N$-values such as $N \lsim 5$, wherever both 
are known. However, only in the DIS case do the $1/N$ terms constitute a 
phenomelogically significant correction over a wide range of moments.

The main application of the present results and, hopefully, their future 
extensions in a more deductive approach -- we note that also an extension
of Ref.~\cite{Mankiewicz:1997gz} to the next-to-leading large-$\nf$ terms
would provide very useful information in the present context -- 
may be in connection with future higher-order diagram calculations, e.g., of 
the fourth-order DIS coefficient functions: 
Firstly they can serve as important checks of such computations which will be 
of unprecedented complexity. Secondly, they will be very useful in combination 
with future partial results such as a fourth-order extension of the fixed-$N$ 
calculations of Refs.~\cite{3loopN}, as fewer computationally costly moments 
will be required for useful $x$-space approximations along the lines of, e.g., 
Ref.~\cite{NV3}.
%
%
\subsection*{Acknowledgments}
S.M. acknowledges support by the Helmholtz Gemeinschaft under contract 
VH-NG-105 and in part by the Deutsche Forschungsgemeinschaft in 
Sonderforschungs\-be\-reich/Transregio~9. 
The research of A.V. has been supported by the UK Science \& Technology 
Facilities Council (STFC) under grant numbers PP/E007414/1 and ST/G00062X/1.
 
{\footnotesize
\setlength{\baselineskip}{0.5cm}

}

\end{document}